\documentclass[preprint,authoryear,round]{imsart}

\let\proof\relax

\expandafter\let\csname equation*\endcsname\relax
\expandafter\let\csname endequation*\endcsname\relax

\RequirePackage[OT1]{fontenc}
\RequirePackage{amsthm,amsmath,enumerate}
\RequirePackage[colorlinks,citecolor=blue,urlcolor=blue]{hyperref}
\usepackage{amssymb}
\usepackage{graphicx,caption,subcaption}
\usepackage{url,color}
\usepackage{epsfig}
\usepackage{amsmath,amssymb,amsthm,mathtools,array,delarray}
\usepackage{natbib}
\usepackage{verbatim}
\usepackage{marginnote}
\usepackage[normalem]{ulem}
\usepackage{xifthen}
\usepackage{float}
\usepackage{tikz}
\usepackage[ruled, % to add horizontal lines. 
lined, %boxed, 
commentsnumbered]{algorithm2e}

\newtheorem{definition}{Definition}
\newtheorem{theorem}{Theorem}

\newtheorem{proposition}{Proposition}
\newtheorem{corollary}{Corollary}
\newtheorem{remark}{Remark}

\usepackage{geometry}

\newcommand{\G}{\mathcal G}

\newcommand{\Gnk}{\G_{n,k}}
\begin{document}

\title{DERGMs: degeneracy-restricted exponential family random graph models}
\runtitle{Degeneracy-restricted ERGMs}

\author{
	Vishesh Karwa\thanks{Karwa was partially supported by NSF TRIPODS+X grant number 1947919.},\\%Work done when the author was a Post Doctoral Fellow at Department of Statistics, Harvard University. Department of Statistics, The Ohio State University
	Temple University\\
	\and \\
	Sonja  Petrovi\'c\thanks{SP is partially supported by the Simons Foundation's the Collaboration Grant for Mathematicians 854770. This work was initially supported by U.S. Air Force Office of Scientific Research  Grant \#FA9550-14-1-0141 to Illinois Tech. A small subset  of the simulations for this work were completed on Illinois Tech's Karlin cluster.}, 
	 Denis Baji\'c\\
	Illinois Institute of Technology
%\email{vishesh@temple.edu,sonja.petrovic@iit.edu, dbajic@hawk.iit.edu}
}
\runauthor{Baji\'c, Karwa, Petrovi\'c}

%\label{firstpage}

\begin{abstract}
	
Exponential random graph models, or ERGMs, are a flexible and general class of models for modeling dependent data. While the early literature has shown them to be powerful in capturing many network features of interest, recent work highlights difficulties related to the models' ill behavior, such as most of the probability mass being concentrated on a very small subset of the parameter space. This behavior limits both the applicability of an ERGM as a model for real data and inference and parameter estimation via the usual Markov chain Monte Carlo algorithms.  
	
To address this problem, we propose a new exponential family of models for random graphs that build on the standard ERGM framework. Specifically, we solve the problem of computational intractability and `degenerate' model behavior by an interpretable support restriction. We introduce a new parameter based on the graph-theoretic notion of degeneracy, a measure of sparsity whose value is commonly low in real-worlds networks. The new model family is supported on the sample space of graphs with bounded degeneracy and is called degeneracy-restricted ERGMs, or DERGMs for short.  Since DERGMs generalize ERGMs -- the latter is obtained from the former by setting the degeneracy parameter to be maximal -- they inherit good theoretical properties, while at the same time place their mass more uniformly over realistic graphs. 
 The support restriction allows the use of new (and fast) Monte Carlo methods for inference, thus making the models scalable and computationally tractable. We study  various theoretical properties of DERGMs and illustrate how the support restriction improves the model behavior. We also present a fast Monte Carlo algorithm for parameter estimation that avoids many issues faced by Markov Chain Monte Carlo algorithms used for inference in ERGMs.   
\end{abstract}

\maketitle		

\section{Introduction} 

Exponential family random graph models, also known as ERGMs for short, are known to be a theoretically flexible class for modeling real world networks. 
There is a growing literature in applications such as \cite{snijders2006new}, \cite{saul2007exploring} and \cite{goodreau2009birds}, 
but also a growing set of contributions on concerns regarding  model complexity and degenerate behavior.  Among the many contributions, we single out recent work by \cite{RinaldoExtremalEdgeTriangle}, \cite{ChatDiacERGMs13}, \cite{ERGMsAreHard}, where various issues of ERGMs have been pointed out and addressed theoretically. While some ERGMs may, as some like to phrase it, `behave badly', this literature also suggests that if we understand this bad behavior, we can still work with this model family - a desirable outcome as the family is quite flexible and broadly encompassing.

Degenerate behavior of some models in the ERGM family that go beyond dyadic independence, as explained in \cite{Handcock03assessingdegeneracy} and, more recently, in \cite{RinaldoFienbergZhu09}, stems from two main issues: The first issue is that given a fixed parameter value, a ``degenerate'' model places most of the probability mass on a small region of the support. 
The second issue is that the subset of parameters where this behavior does \emph{not} happen can be very small. 
This property is then naturally implicated in other problems such as estimation, in particular, non-convergence of MCMC-MLE estimates.  
A popular algorithm for estimation is to %Estimation is done by approximating 
approximate the log likelihood using importance sampling from the model with a fixed parameter $\theta_0$, usually via an MCMC sampler. To obtain an accurate approximation of the log likelihood, the standard MCMC sampler must generate samples from the region where the mass is concentrated. Since the mass is tightly concentrated on a small region, the MCMC sampler must start with a parameter very close to MLE, otherwise estimation fails.  See \cite{snijders2002markov} for the Robbins-Monro algorithm, which need not start with a parameter close to the true MLE for the estimation to not fail.

The literature offers several approaches to address the issue of model degeneracy, including the study of curved ERGMs with alternating $k$-star and $k$-triangle terms and geometrically weighted edge wise shared partner terms  (\cite{snijders2006new}, \cite{HunterandHandcock}, \cite{hunter2008goodness}); dyad-independent ERGMs (ERGMs that assume the dyads are independent) with sparsity assumptions (\cite{krivitsky2011adjusting}, \cite{kolaczyk2015question});  ERGMs with local dependence (\cite{schweinberger2015local}),  nonparametric ERGMs (\cite{thiemichen2017stable}); and an example of a re-parametrized  ERGM that appears in  \cite{horvat2015reducing},  who study the  edge-triangle ERGM and propose a one-to-one transformation of the sample space that renders the model non-degenerate.

Our work contributes to this understanding and proposes a natural support restriction of ERGMs to \emph{sparse graphs} and without the dyadic independence assumption. The class of sparse graphs that we consider are called $k$-degenerate graphs, defined below. We show that restricting support to $k$-degenerate graphs provably reduces the degenerate behavior.  %and is interpretable in applications. 
To formally show improvement in model behavior, we rely on the notion of model degeneracy and stability as defined in \cite{schweinberger2011instability} as our starting points. Schweinberger defined stability of sufficient statistics and showed that instability leads to model degeneracy. We generalize and strengthen this definition to support-restricted ERGMs, including DERGMs, and prove that stability implies non-degeneracy of the model.

To decide how to restrict support, we build our intuition on the observation that has been noted in much of the network literature: many real-world networks are sparse in some sense. While there are many different notions of sparsity, we use the following class of sparse graphs: a network is said to be sparse if it has bounded \emph{degeneracy}\footnote{Sadly, the two fields - graph theory and statistics - use the same term, degeneracy, for two different concepts. We will show that degeneracy-restricted graphs lead to non-degenerate models.}, 
defined as follows (see Remark~\ref{rmk:discussionOfDegeneracy} for equivalent descriptions). 
\begin{definition}[Degeneracy of a graph $g$, \cite{lick1970k, Seidman83}]

	 The \emph{k-core} $H_k(g)$ of  $g$ is the maximal subgraph of $g$ in which every vertex has degree at least $k$. Here, maximal means with respect to inclusion.  
	 The \emph{degeneracy} of a graph $g$ is the maximum index of its non-empty core: $\max\{k: H_k(g)\neq\emptyset\}$. 
 \end{definition}

\paragraph{Examples:}
	Consider a star graph on $n$ nodes. It has degeneracy $1$. On the other extreme, a fully connected graph has degeneracy $n$. Note that the degree of the star graph is $n-1$, but its degeneracy is $1$. Figure \ref{fig:core-example} shows a more interesting example of a small network with degeneracy 4, along with its cores.

\begin{figure}[h]
	\begin{tikzpicture}
	[scale=.2,auto=center,every node/.style={circle,fill=black},inner sep=1.5pt]
	\node (n1) at (-9,0)  {}; %L
	\node (n2) at (5.5,0)  {};
	\node (n3) at (4,3)  {};
	\node (n4) at (8.5,0)  {};
	\node (n5) at (10,3)  {};
	\node (n6) at (5.5,6)  {};
	\node (n7) at (0,0)  {}; %L
	\node (n8) at (-6,0)  {}; %L
	\node (n9) at (-3,2)  {}; %L
	\node (n10) at (0,6)  {}; %M
	\node (n11) at (-2.121312,8.12132)  {}; %M
	\node (n12) at (-3,6)  {};
	\node (n13) at (0,9)  {};
	\node (n14) at (-2,3.5)  {};
	\node (n15) at (8.5,6)  {};
	\node (n16) at (-1.5,-2.5)  {}; %L
	\node (n17) at (-4.5,-2.5)  {}; %L
	\foreach \from/\to/\weight in {
		n15/n3/1, n15/n4/1, n15/n5/1, n15/n6/1, n2/n6/1, n3/n6/1, n5/n6/1,
		n2/n5/1, n2/n3/1, n3/n4/1, n2/n4/1, n3/n5/1, n4/n5/1,
		n3/n10/1,n10/n11/1,n10/n12/1,n10/n13/1,
		n7/n3/1,
		n1/n8/1, n7/n8/1, n7/n17/1,  n8/n9/1,n7/n9/1,  n7/n16/1,  n8/n17/1, n16/n17/1,
		n16/n2/1 }
	\draw (\from) --(\to);
	\end{tikzpicture}
	\quad\quad\quad\quad\quad
	\begin{tikzpicture}
	[scale=.2,auto=center,every node/.style={circle,fill=black},inner sep=1.5pt]
	\node (n2) at (5.5,0)  {};
	\node (n3) at (4,3)  {};
	\node (n4) at (8.5,0)  {};
	\node (n5) at (10,3)  {};
	\node (n6) at (5.5,6)  {};
	\node (n7) at (0,0)  {}; %L
	\node (n8) at (-6,0)  {}; %L
	\node (n9) at (-3,2)  {}; %L
	\node (n15) at (8.5,6)  {};
	\node (n16) at (-1.5,-2.5)  {}; %L
	\node (n17) at (-4.5,-2.5)  {}; %L
	\foreach \from/\to/\weight in {
		n15/n3/1, n15/n4/1, n15/n5/1, n15/n6/1, n2/n6/1, n3/n6/1, n5/n6/1,
		n2/n5/1, n2/n3/1, n3/n4/1, n2/n4/1, n3/n5/1, n4/n5/1,
		n7/n3/1,
		n7/n8/1, n7/n17/1,  n8/n9/1, n7/n9/1,  n7/n16/1,  n8/n17/1, n16/n17/1,
		n16/n2/1 }
	\draw (\from) --(\to);
	\end{tikzpicture}
	\quad\quad\quad\quad\quad
	\begin{tikzpicture}
	[scale=.2,auto=center,every node/.style={circle,fill=black},inner sep=1.5pt]
	\path [use as bounding box,red] (4,-2.5) rectangle (9,8.5);
	\node (n2) at (5.5,0)  {};
	\node (n3) at (4,3)  {};
	\node (n4) at (8.5,0)  {};
	\node (n5) at (10,3)  {};
	\node (n6) at (5.5,6)  {};
	\node (n15) at (8.5,6)  {};
	\foreach \from/\to/\weight in {
		n15/n3/1, n15/n4/1, n15/n5/1, n15/n6/1, n2/n6/1, n3/n6/1, n5/n6/1,
		n2/n5/1, n2/n3/1, n3/n4/1, n2/n4/1, n3/n5/1, n4/n5/1 }
	\draw (\from) --(\to);
	\end{tikzpicture}		\caption{An example of a small graph $g$ (left), its $2$-core (center), and its 3- and 4-core (right). Adapted from \cite{KarPelPetStaWilb:shellErgm}}
	\label{fig:core-example}
\end{figure}
Many real world networks tend to have small degeneracy with respect to the number of nodes. The table below (adapted from \cite{KarPelPetStaWilb:shellErgm}) shows examples of some sample networks whose degeneracy is much less compared to the number of nodes.
\begin{table}[ht]
	\centering
	\begin{tabular}{l lll}
		Network Dataset \qquad \qquad & Nodes \qquad   \qquad  &Edges   \qquad	 \qquad  & Degeneracy\\ 
		\hline
		Scotland  		& 244	& 256	&  4 \\ 
		Geom 			& 7343	& 11898	& 21 \\ 
		NDyeast 		& 2114	& 2277	&  5 \\ 
		NetScience 		& 1589	& 2742	& 19 \\
		USpowerGrid		& 4941	& 6594	&  5 \\
		Erd\H{o}s		& 6927	& 11850	& 10 \\ 
		\hline
	\end{tabular}
\end{table}

 Without further ado, let us define the model class, and then discuss the graph-theoretic notion more intuitively.

Let $\G_n$ be the set of all simple graphs on $n$ nodes. This sample space definition for ERGMs is standard, though extensions exist to valued graphs, see \cite{valuedERGM}. Recall that the ERGM with sufficient statistics vector $t=(t_1,\dots,t_d)$ defined on the parameter space $\Theta\subset \mathbb R^d$  places the following probability on any $g\in\G_n$: 
\begin{equation} \label{eq:ERGM}
P_{ERGM}(G=g)=\frac{\exp\{\theta^T\cdot t(g)\}}{c(\theta)},
\end{equation}
where $\theta=(\theta_1,\dots,\theta_d)$ are the canonical parameters, $c(\theta)$ is the normalizing constant $c(\theta)=\sum_{g\in\G_n}\exp\{\theta^T\cdot t(g)\}$, and the set of possible parameters is given by $\Theta = \{\theta \in \mathbb R^d: c(\theta) < \infty \}$. 
In the corresponding DERGM, we simply restrict the support of the model from $\G_n$ to the set of all graphs on $n$ nodes whose degeneracy is at most $k$. 

\begin{definition}[DERGM]
	\label{defn:OurModel}
	Denote by $\Gnk$ the set of all graphs on $n$ nodes whose degeneracy is at most $k$. Choose a vector of graph statistics $t=(t_1,\dots,t_d)$.  The \emph{degeneracy-restricted exponential random graph model}, or DERGM for short,  with sufficient statistics vector $t$ places the following probability on a graph on $n$ nodes: 
	\begin{equation} \label{eq:DERGM}
	P_{DERGM}(G=g)= \begin{cases}
	\exp\{\theta^T\cdot t(g)\}\cdot c_k(\theta)^{-1}, & \quad \text{if } g\in\Gnk \\
	0, &\quad \text{otherwise},
	\end{cases}
	\end{equation}
	where $ c_k(\theta)$ is the \emph{modified} normalizing constant \[  c_k(\theta)=\sum_{g\in\Gnk} \exp\{\theta^T\cdot t(g)\},\] 
	and the set of possible parameters is given by
	 \[\Theta = \{\theta \in \mathbb R^d: c_k(\theta) < \infty \}.\]
\end{definition}
\noindent Note that setting $k=n-1$ reduces the DERGM to the usual ERGM. %, thus this is a more general model class.  

Section~\ref{sec:simulations} illustrates the effect of changing the degeneracy parameter value on the model behavior. 
For example, following \cite{schweinberger2011instability}, we investigate whether models exhibit  excessive sensitivity, where  small changes in the values of the natural parameters lead to large changes in the mean-value parameter and show an example where DERGMs do not exhibit such excessive sensitivity when compared to the corresponding ERGM. 
In addition, simulation results in Section~\ref{sec:simulationsFitting}  provide evidence that the
 parameter estimates of a DERGM are not too different from the corresponding ERGM, in cases where both can be estimated.  
That is, even if the true data generating distribution is an ERGM, there is very little or no difference in fitting a DERGM. 

One may ask, what is the point of fitting a DERGM in such cases when the ERGM parameters can also be estimated? Our reasoning is that in such cases, one may think of support restriction as a means of improving the properties of the MCMC-MLE estimation procedure by preventing the Markov chain from visiting states that are extremal (e.g. graphs that are complete or near complete). Moreover, we believe that any reasonable ERGM that fits a real world data will place very little mass on graphs with large degeneracy (this can be demonstrated by fitting an ERGM, simulating a lot of graphs from the ERGM and recording the degeneracy parameter). Further, these experiments show that in cases where ERGMs cannot be fit, fitting a DERGM will give us reasonable parameter estimates.

\begin{remark}\rm
\label{rmk:discussionOfDegeneracy}
 Graph degeneracy has  other characterizations; for instance, a $k$-degenerate graph admits an ordering of its vertices $v_1, \ldots, v_n$ such that vertex $v_i$ has at most $k$ neighbors after it in the ordering; thus a bounded-degeneracy  graph means there exists a vertex with  few neighbors. In fact, another characterization is that in a \emph{$k$-degenerate} graph, every induced subgraph has a vertex of degree at most $k$.  Hence,  bounding the degeneracy of a graph is a \emph{weaker constraint} than bounding the overall node degree in the graph, and it is also weaker than bounding the so-called $h$-index, which means that  most nodes have few neighbors. 
 For supporting evidence of low-degeneracy network data, see \cite[Section 3.1]{KarPelPetStaWilb:shellErgm}, where the authors compute degeneracy of each of the undirected graphs in the \cite{Pajek} database. 
A secondary reason to consider this support restriction is that restricting to bounded-degeneracy graphs makes many sub graph counting algorithms computationally efficient: for example,  all the maximal cliques can be enumerated in  polynomial time in the case of bounded degeneracy, while in general the problem is NP-hard.  
\end{remark}
\begin{remark}\rm
	We want to emphasize the fact that bounding the degeneracy of a graph does not impose any bound on the maximum degree. Consider, for example, a star graph on $n$ nodes. The maximum degree is $n-1$, but the degeneracy is only $1$. In fact, the key reason for bounding the degeneracy and not the degree is that one gets a class of graphs that can have very high degree nodes, but are still sparse in some sense. 
\end{remark}

\begin{remark}\rm
\label{rmk:whyKobserved}
A discussion on the choice of $k$ is in order.  
The problem of simultaneously  estimating $\theta$ and $k$ from $g_{obs}$ seems quite difficult, since changing $k$ changes the support of the model.
 We consider the choice of $k$ akin to the problem of model selection, as different values of $k$ describe different models. Valid choices of $k$ range from the observed value $k_{obs}$ to $n-1$, where $k=n-1$ reduces to the usual ERGM.
 Setting $k = k_{obs}$ seems to be a reasonable choice (and it is the minimal choice, otherwise the model places $0$ probability on the observed graph), for now, given that in most real world networks $k_{obs}$ is much smaller than $n$. More importantly, we will show in Section \ref{sec:stability} that setting $k \ll  n$ leads to improved model behavior, and in addition we prove a lower bound on the size of the support of such a DERGM compared to the full ERGM.  Choosing smaller values of $k$  leads to a likelihood function that is better behaved, eliminates dense graphs from the support, and reduces model degeneracy. We show this in detail theoretically and by simulations. 
\end{remark}

A summary of the contributions of the remainder of this manuscript is as follows. 
In Section~\ref{sec:stability}, we prove that the support of a DERGM with $k\ll n$ is not too small compared to $k=n-1$, extend and strengthen the definition of stability of sufficient statistics from \cite{schweinberger2011instability}, and prove that stability implies that the DERGM is non-degenerate. We also present an example of an unstable ERGM whose counterpart DERGM is stable, namely, one with a two-dimensional parameter space whose sufficient statistics are the number of edges and number of triangles in the graph. The degeneracy of the edge-triangle model is studied in detail by \cite{RinaldoFienbergZhu09}. 
In Section~\ref{sec:MLEofDERGM} we discuss the general estimation problem in DERGMs and address various aspects of the problem, including existence of the MLE and approximate MLE.  Section~\ref{sec:modelSampler} also provides a straightforward Metropolis-Hastings algorithm to sample from the model. 
In Section~\ref{sec:simulationsOnModelBehavior} we provide simulation results that support the theoretical claims about degeneracy-restricted ERGMs. Specifically, we discuss the choice of $k$, why DERGMs do not suffer from the same estimation issues that arise  in standard ERGMs, model degeneracy issues and how they disappear for smaller values of $k$. We focus on the edge-triangle models as the running example; these  are well-studied sufficient statistics that arise naturally when considering Markov dependence, see for example \cite{frank1986markov} and recent complementary work \cite{LauritzenRinaldoSadeghi2018exchNtwk}. As a running example in \cite{RinaldoFienbergZhu09}, it is also the natural example to compare ERGM behavior to DERGMs. 
Section~\ref{sec:simulationsFitting} includes simulation studies on real-world network data, including those where a DERGM fits but ERGM fails to converge, as well as examples where both models fit. 
Section~\ref{sec:uniformSampler} derives uniform samplers of the sample space $\Gnk$ --- which were used throughout Section~\ref{sec:simulationsOnModelBehavior} ---  and further discusses some of the algorithmic considerations pertaining to scalability and applicability. 
The {\tt R} and {\tt Python} code used to run the simulations in Section~\ref{sec:simulations}, along with implementations of the main algorithms from Section~\ref{sec:uniformSampler}, is available on {\tt GitHub} under \cite{dergmGitHub}.

%%%%%%=====================%%%%%%

\section{Non-degeneracy and Stability  of DERGMs}\label{sec:stability}

In this section, we formally show that restricting the support of an ERGMs to $k$-degenerate graphs improves model behavior.  \cite{schweinberger2011instability} showed that the degenerate behavior of an ERGM is closely tied with the notion of ``stability'' of sufficient statistics that are used to define the ERGM. In particular,  ``un-stable'' sufficient statistics lead to excessive sensitivity of the model, which in turn leads to degenerate model behavior and impacts the MCMC-MLE estimation. 
We extend the notion of stability to support-restricted models and tie it to the support size of a model. 
 Roughly, a sufficient statistic is stable if it can be strictly upper-bounded by the log of support size of the model. In an ERGM, the log of support size is of order $O(n^2)$ and hence any sufficient statistic that grows faster than $O(n^2)$ is considered unstable. This includes the number of triangles and number of two-stars, both of which grow at a rate of $O(n^3)$. This unstable behavior leads to excessive sensitivity and  degeneracy of the edge-triangle ERGM. 
 DERGMs, on the other hand, are defined by restricting the support size and include only $k$-degenerate graphs for a fixed $k$. Restricting the support to $k$-degenerate graphs induces stability of sufficient statistics such as triangles and two-stars, which in turn improves model behavior.   Furthermore, if $k$ is fixed, the number of edges and triangles is of the same order, so the triangle term cannot dominate the edge term; see Proposition~\ref{prop:stableSufficientStatistics}.

First, we study the size of the support of DERGMs in Theorem~\ref{thm:lowerboundSupport}, generalize the notion of  stable sufficient statistics in Definition \ref{def:stablestats}, and show stability holds for the edge-triangle DERGM in Proposition~\ref{prop:stableSufficientStatistics} (\cite{schweinberger2011instability} showed the edge-triangle ERGM is unstable; cf.\ \cite{RinaldoFienbergZhu09}). 
Then, in Theorem \ref{thm:non-degenerate}, we show that  any DERGM with stable sufficient statistics is not degenerate under the formal definition of asymptotic non-degeneracy from \cite{schweinberger2011instability}. 
\paragraph{Order notation.}
Many of the results in this paper are asymptotic and use the order notation. For readers' convenience, we include the definitions we use: the `big-O', denoted by $O(\cdot)$;  `little-O', or $o(\cdot)$;  `big-Omega', $\Omega(\cdot)$; and `Theta', $\Theta(\cdot)$.  They offer convenient shorthand for comparing the asymptotic growth of two functions $f(n)$ and $g(n)$, $n\in\mathbb Z_{\geq 0}$:  
\begin{enumerate}
	\item   $f(n)$ is $O(g(n))$ if there exists a constant $c>0$ and an integer $n_0$, such that for all $n > n_0$, the bound $f(n) \leq c \cdot g(n)$ holds. 
	\item  $f(n)$ is $o(g(n)))$ if for all constants $c>0$, there exists an integer $n_0$ such that for every $n \geq n_0$, $f(n) < c \cdot g(n)$.
	\item  $f(n)$ is $\Omega(g(n))$ if there exists a constant $c>0$ and an integer $n_0$, such that for all $n > n_0$, such that $f(n) \geq c \cdot g(n)$.
	\item   $f(n)$ is $\Theta(g(n))$ if $f(n)$ is $O(g(n))$ and $f(n)$ is $o(g(n))$.
\end{enumerate}
\subsection{Support size of DERGMs} 
The number of graphs in the support of a ERGM is $2^{{n \choose 2}}$. Since a DERGM restricts the support, a natural question that arises is: what is the number of graphs in the support of a DERGM with degeneracy parameter $k$?  Unfortunately, there are no simple formulas to count the number of $k$-degenerate graphs; nonetheless, we can obtain an asymptotic lower bound as follows. % on the number of graphs in the support of a DERGM. 

\begin{theorem}[Support size of DERGMs]
	\label{thm:lowerboundSupport}
	Let $S_k(n)$ denote the number of simple graphs with $n$ nodes and degeneracy at most $k$. Then, for a fixed $k$, there exist positive constants $c_1,c_2 > 0$ and an integer $n_0$ such that for all $n > n_0$, 
	$$c_1 \cdot n \log n \leq \log S_k(n) \leq c_2 \cdot n \log n$$
	That is, for a fixed $k$, and as $n$ goes to infinity, $\log S_k(n) = \Theta\left( n \log n \right).$
	On the other hand, for $k=n-1$, $\log S_{n-1}(n) = \Theta(n^2).$
\end{theorem}
Theorem \ref{thm:lowerboundSupport} is an asymptotic statement that gives an asymptotic upper and lower bound on the support size of DERGMs, when $k$ is a fixed constant. For the finite sample settings, we  can consider $k = O(1)$, i.e. % \sout{as a small fixed constant} 
 $k$ is a bounded from above by a constant, whereas $n$ is increasing. (As a practical example, $n$ may be $5000$, but $k$ may be $50$ or even $10$.) Under such settings, Theorem 1 shows that there are about $O(2^{n\log n})$ graphs in the support of DERGM. On the other hand, the ERGM has $O(2^{n^2})$ graphs. Note that $S_{n-1}(n)$ is the size of the support of the full ERGM. 
We found two interesting properties: that parameter estimates of a DERGM do not change drastically from that of the corresponding ERGM, see Section \ref{sec:ergm&dergmfits} for a concrete example; and that the graphs eliminated from the support of the ERGM are precisely the ones that cause instability issues, as illustrated in the next result. 

\begin{proof}[Proof of Theorem~\ref{thm:lowerboundSupport}]
	We derive both upper and lower bounds for the DERGMs support size. A natural lower bound on the number of $k$-degenerate graphs is the number of \textit{well-ordered} $k$-degenerate graphs. A \textit{well-ordered} $k$-degenerate graph is a labeled graph with vertex-labels $1, \ldots, n$ such that the
	ordering of the vertices by their labels is a well-ordering of the graph.
	From \cite{BauerEtAlSamplingDegenerate}, the number of \emph{well-ordered} graphs with degeneracy at most $k$ is given by
	$$ D_k(n) = D_k(n-1) \cdot   \sum_{i=0}^{\min(n-1,k)} {n-1 \choose i}.$$ 
	By definition, $D_k(n)$ is a lower bound on the $S_k(n)$. % , i.e. $S_k(n) \geq D_k(n)$.
	Applying the recursion, for a constant $k$, we get
	$$D_k(n) = \left( \sum_{i=0}^{k} {n-1 \choose i} \right) \cdot  \left( \sum_{i=0}^{k} {n-2 \choose i} \right) \ldots \cdot \left( \sum_{i=0}^{k} {k \choose i} \right) \cdot \left( \sum_{i=0}^{k-1} {k-1 \choose i} \right) \cdot \left( \sum_{i=0}^{1} {1 \choose i} \right),$$
	which further simplifies as follows:
	\begin{align*}
	D_k (n)
	&= \prod_{r=k+1}^{n-1} \sum_{i=0}^{k} {r \choose i} \cdot \prod_{r=1}^{k}\sum_{i=0}^{r} {r \choose i}\\
	&=\prod_{r=k+1}^{n-1} \sum_{i=0}^{k} {r \choose i} \cdot \prod_{r=1}^{k}2^r\\
	&=\prod_{r=k+1}^{n-1} \sum_{i=0}^{k} {r \choose i} \cdot 2^{{k \choose 2}}.
	\end{align*}
	Taking logarithms gives 
	\begin{align*}
	\log D_k (n)
	&=\sum_{r=k+1}^{n-1} \log \left(\sum_{i=0}^{k} {r \choose i} \right) + {k \choose 2}\log 2\\
	&\geq \sum_{r=k+1}^{n-1} \log {r \choose k} + {k \choose 2}\log 2.
	\end{align*}
    Note that the second term depends only on $k$ and hence we can focus on the first term. Let
	$$T_k(n) := \sum_{r=k+1}^{n-1} \log  {r \choose k}. $$
	Using the lower bound ${r \choose k} \geq (r/k)^k$, we get,
	\begin{align*}
	T_k(n) &\geq k \cdot \sum_{r=k+1}^{n-1} \log (r/k) \\
	&\geq k \cdot \left(\sum_{r=k+1}^{n-1} \log r\right ) - k \log k (n-k-1)\\
	&=k \cdot \left(\sum_{r=1}^{n-1} \log r - \sum_{r=1}^k \log r\right )  - k\log k (n-k-1) \\
	&= k \cdot \left(\log (n-1)! -  \log k! \right )  - k\log k (n-k-1) \\
	&=\Omega(n \log n).
	\end{align*}
	Thus the claimed lower bound follows: $\log S_k(n) \geq \log D_k(n) \geq T_k(n) = \Omega(n \log n)$.
	
For the upper bound on the support size of $k$-degenerate graphs, we will use the following strategy. Let $\#G(n,\leq m)$ denote the number of graphs on $n$ nodes with \textit{at most} $m$ edges,  we will show  below that
\begin{align}
\label{atmostm}
\log \#G(n,\leq m) \leq 2m \cdot \log (en)
\end{align}
From Proposition \ref{prop:stableSufficientStatistics} below, the maximum number of edges in a $k$-degenerate graph is $k \cdot n - {(k+1) \choose 2}$. Using the fact that 
$$\Gnk \subset 
G(n,\leq m),$$ where $m =k \cdot n - {(k+1) \choose 2}$, we have the following upper bound:
\begin{align*}
\log S_k(n) &\leq \log \#G\left(n,\leq k \cdot n - {(k+1) \choose 2}\right) \\
& \leq 2 \left(k \cdot n - {(k+1) \choose 2}\right) \log (en) \\
& < 2k \cdot n \log (en) = O(n \log n)
\end{align*}
Finally, to see that the upper and lower bounds for the case when $k=n-1$ hold, note that $k=n-1$ is the full ERGM and we have $2^{n \choose 2}$ graphs in the support of an ERGM. Thus $\log S_{n-1}(n) = \log 2^{n \choose 2} = \Theta(n^2)$.

All that remains to be shown is equation \ref{atmostm}. Note that the number of graphs on $n$ nodes with $m$ edges is ${{n \choose 2} \choose m}$, since there are ${n \choose 2}$ possible locations to choose from and place the $m$ edges. Now the number of graphs with at most $m$ edges is given by 
\begin{align*}
\#G(n,\leq m) &= \sum_{i=0}^{m} {{n \choose 2} \choose i} \leq \left(\frac{e{n \choose 2}}{m}\right)^m, 
\end{align*}
from the well known fact $\sum_{i=0}^m {n \choose i} \leq \left(\frac{en}{m}\right)^m$.
Taking logs, we get
\begin{align*}
\log \#G(n,\leq m) &\leq \log \left(\frac{e{n \choose 2}}{m}\right)^m \\
&\leq \log \left(\frac{en^2}{m}\right)^m \\
&\leq 2m\log en.
\end{align*}
 
\end{proof}

\subsection{Stability of Sufficient Statistics}
By restricting the support to include only those graphs with degeneracy at most $k$, where $k$ is small compared to $n$, we eliminate ``dense'' graphs from the model. In turn, this  has a stabilizing effect on the sufficient statistics. A formal definition of a stable sufficient statistic in ERGMs is given in \cite{schweinberger2011instability}.  %We generalize that definition so that it can be applied to models with support restriction, and strengthen it slightly. 

\begin{definition}[Stable sufficient statistics]
	\label{def:stablestats}
Let $S_k(n)$ be the size of support of a DERGM with sufficient statistic $t(g)$. Then $t(g)$ is said to be \emph{stable} if for any constant $C > 0$ there exists an integer $n_0$ such that for every $n \geq n_0$
$$ \underset{g \in \mathcal{G}_{n,k}}{\max} t(g) < C \cdot \log S_k(n) $$
or in other words, $\underset{g \in \mathcal{G}_{n,k}}{\max} t(g) \in o(\log S_k(n))$.
On the other hand $t(g)$ is said to be unstable if for any $C > 0$, however large, 
$$ \underset{g \in \mathcal{G}_{n,k}}{\max} t(g) \geq C \cdot \log S_k(n) $$  
A vector of sufficient statistics is stable if all the components of the vector are stable, if any component is unstable, the vector of sufficient statistics is unstable. 
\end{definition}

Roughly, a sufficient statistic is stable if it can eventually be strictly upper-bounded by the log of the support size of the DERGM. If it cannot be upper bounded by the log of support size, then  it is unstable. For an ERGM, with no support restriction, this definition reduces to strictly upper bounding the sufficient statistic by ${n \choose 2}$, where $n$ is the number of nodes and it strengthens the definition of stable sufficient statistics in \cite{schweinberger2011instability}. The edge-triangle ERGM is not stable due to the instability of the number of triangles, as shown in \cite{schweinberger2011instability}. However, it turns out that the edge-triangle DERGM is stable. % as long as $k \ll  n$.

\begin{proposition}
	\label{prop:stableSufficientStatistics}
	Let $e(g)$ be the number of edges and $\triangle(g)$ be the number of triangles in a graph. Then
	\begin{enumerate}
		\item $\underset{g \in \mathcal{G}_{n,k}}{\max} e(g) = k \cdot n - {(k+1) \choose 2}$
		\item $\underset{g \in \mathcal{G}_{n,k}}{\max} \triangle(g) = {k \choose 3} + {k \choose 2}(n - k).$
	\end{enumerate}
\end{proposition}
\begin{proof}
	For this proof, we use the notion of a \emph{shell index} of a node: define the {$i$-th shell}  of a graph $g$ to be the difference of the two consecutive cores $H_i(g)\setminus H_{i-1}(g)$. Note that a node may belong to more than one core, but shell membership is unique. Thus we say that a vertex $v$ is said to have \emph{shell index} $i$ if  $v\in H_i(g)$ but $v\not\in H_{i+1}(g)$.

 	For any given network, the shell sequence $s_1 \leq s_2 \ldots \leq s_n$ is the sorted sequence of shell indices of each node. From Proposition 10 in \cite{KarPelPetStaWilb:shellErgm}, the maximum number of edges in a graph with a shell sequence $s_1 \leq s_2 \ldots \leq s_n$ is given by:
	$$ {k \choose 2} + \sum_{i=1}^{n-k} s_i.$$
	This expression is maximized by graphs in which all the nodes are in the $k^{th}$ core, which  has a shell sequence $s_1 = k, s_2 = k, \ldots s_n = k$. Thus the maximum number of edges in a $k$-degenerate graph is
	$$ {k \choose 2} + \sum_{i=1}^{n-k} k = \frac{k(k-1)}{2} + k(n-k) = nk - {(k+1) \choose 2}.$$
	
	Similarly, from Proposition 12 in \cite{KarPelPetStaWilb:shellErgm}, the maximum number of triangles in a graph with shell sequence $s_1 \leq s_2 \ldots \leq s_n$ is given by:
	$$ {k \choose 3} + \sum_{i=1}^{n-k} {s_i \choose 2}.$$
	This expression is maximized also  when all the nodes are in the $k^{th}$ core. Thus the maximum number of triangles is 
	$$ {k \choose 3} + \sum_{i=1}^{n-k} {k \choose 2} = {k \choose 3} + (n-k) {k \choose 2}. $$

\end{proof}

Proposition \ref{prop:stableSufficientStatistics} shows that the number of triangles in a $k$-degenerate graph is $O(n)$, whenever $k = O(1) $. (In fact $k$ can be allowed to grow with $n$, albeit slowly, see the next theorem) On the other hand, without any restriction on the degeneracy, the number of triangles can be as large as $O(n^3)$ making the ERGMs unstable.  The number of triangles in $k$-degenerate graphs is linear in $n$, which make them a good candidate to model sparse graphs, which are commonplace in the real world. 

In Theorem \ref{thm:stable edge-triangle}, we use Proposition \ref{prop:stableSufficientStatistics} to show that the edge-triangle DERGM is stable. 
The way we defined a DERGM assumes that $k$ is fixed; however, note that Theorem \ref{thm:stable edge-triangle} shows that $k$ can grow with $n$, albeit slowly: For instance, if $k$ grows with $\sqrt{\log(n)}$, then the sufficient statistics are still stable.

\begin{theorem}[Stability of Edge-Triangle DERGM]
	\label{thm:stable edge-triangle}
	Consider the edge-triangle dergm with the vector of sufficient statistics $t(g) = (e(g), \triangle(g))$ where $e(g)$ is the number of edges  and $\triangle(g)$ is the number of triangles. The edge-triangle dergm is stable as long as $k = o(\sqrt{\log n})$.
\end{theorem}
\begin{proof}
	We need to show that for all $c>0$, there exists $n_0$, there exists $n > n_0$ such that $\max_g (e(g), \triangle(g)) < c \cdot \log S_k(n))$ where the max is over the support set $g \in \mathcal{G}_{n,k}$. Fix a $g$ in $\mathcal{G}_{n,k}$. From Proposition \ref{prop:stableSufficientStatistics}, we have,
	\begin{align*}
	(e(g), \triangle(g))  	&\leq  \left(k \cdot n - {(k+1) \choose 2},{k \choose 3} + {k \choose 2}(n - k)\right) \\
										 	&\leq O(k \cdot n, k^2 \cdot n)
	\end{align*}
Thus, if $k = o(\sqrt{\log n})$, we have, $(e(g), \triangle(g)) = o(n\log n) = o(\log S_k(n))$.
\end{proof}

\subsection{Non-degeneracy of DERGMs}
We now show that stability of sufficient statistics implies that a DERGM is non-degenerate. Let us begin by defining degeneracy of a distribution, or more precisely the degeneracy of a parameter associated with a distribution.  Consider a DERGM defined by the parameter vector $\theta$ and sufficient statistics $t(g)$ and let $M_k(\theta)$ be the set of modes, i.e.
$$M_k(\theta)  = \underset{g \in \mathcal{G}_{n,k}}{\arg\max } \frac{e^{\theta^T \cdot t(g)} }{c_k(\theta)}.$$ 
One also defines a set of $\epsilon$-modes for any $0< \epsilon<1$: 
\[
M_{\epsilon,k}(\theta)  = \{ G \in \mathcal{G}_{n,k} : e^{\theta^T \cdot t(G)}  > (1-\epsilon) \underset{g \in \mathcal{G}_{n,k}}{\max } e^{\theta^T \cdot t(g)}\}.
\]
A parameter $\theta$ is said to be \emph{asymptotically degenerate}  if the distribution induced by $\theta$ asymptotically places  all of its mass on its modes. 

\begin{definition}[Asymptotically degenerate parameters, see also \cite{schweinberger2011instability}] 
A parameter $\theta$ is said to be asymptotically degenerate if 
$$\lim\limits_{n \rightarrow \infty} \mathbb P_{\theta}(G \in M_k(\theta)) = 1.$$
\end{definition}
If, on the other hand, 
$\lim\limits_{n \rightarrow \infty}\mathbb P_{\theta}(G \in M_k(\theta)) $ is bounded away from $1$, 
the model is asymptotically non-degenerate. 
We define asymptotic near-degeneracy for DERGMs similarly using $\epsilon$-modes.  

As \cite{schweinberger2011instability} discusses, strict degeneracy in discrete exponential families isn't attainable, thus $\theta$ is said to be near-degenerate if the mass concentrates on $\epsilon$-modes. 
The same reference proves that unstable sufficient statistics lead to near degenerate distributions. 
In the following result we prove that, under a technical condition that the number of graphs in the $\epsilon$-modes grows slower than square root of the model support size,   stability  implies non-(near-)degeneracy in the more general case of DERGMs. 
\begin{theorem}[Stability implies non-(near)-degeneracy]
	\label{thm:non-degenerate}
	Consider any DERGM with parameter vector $\theta$ and the vector of sufficient statistics $t(g)$, and a bounded and fixed degeneracy parameter $k$. 
	Suppose that $t(g)$ is stable. Assume $\theta\in\Theta$ is such that there exists a constant $c$ and an $n_0$ such that for all $n> n_0$, $|M_{\epsilon,k}(\theta)| < c \cdot \sqrt{S_k(n)}$, that is the number of graphs in the set of $\epsilon$ modes does not grow larger than the square root of the total number of graphs in the model support. 
%	as long as $\theta\in\Theta$ is such that not all graphs are the $\epsilon$-modes (in other words, $|\overline{M_{\epsilon,k}(\theta)}| \geq 1$), 
	Then,  the DERGM is asymptotically non-(near)-degenerate at $\theta$. 
\end{theorem}
\begin{proof} 	 
	To show that a DERGM is not near-degenerate, we need to show that  
	$\lim\limits_{n \rightarrow \infty} \mathbb P_{\theta}(G \in M_{\epsilon,k}(\theta))  < 1$.  
	That is, we need to show that for every $0<\epsilon<1$, however small, 
	$\mathbb P_{\theta}(G \in M_{\epsilon,k}(\theta)) $ is bounded away from $1$ asymptotically. 

\begin{align*}
		\mathbb P_{\theta}(G \in M_{\epsilon,k}(\theta)) &= \frac{1}{c_k(\theta)} \sum_{g \in M_{k,\epsilon}(\theta)} \exp(\theta^T\cdot  t(g)) \\
		&= \frac{\sum_{g \in M_{\epsilon,k}(\theta)}{\exp(\theta^T \cdot t(g)) }}{ \sum_{g \in \mathcal{G}_{n,k} }{\exp(\theta^T \cdot t(g)) } }\\
		&= \frac{\sum_{g \in M_{\epsilon,k}(\theta)}{\exp(\theta^T \cdot t(g)) }}{\sum_{g \in M_{\epsilon,k}(\theta)} \exp(\theta^T \cdot t(g)) + \sum_{g \in \mathcal{G}_{n,k} \setminus M_{\epsilon,k}(\theta)} \exp(\theta^T \cdot t(g))} \\
		&=\frac{1}{1 + r_n},
\end{align*} where 
\begin{align*}
		r_n &= \frac{\sum_{g \in \mathcal{G}_{n,k} \setminus M_{\epsilon,k}(\theta)} e^{\theta^T \cdot t(g)}}{\sum_{g \in M_{\epsilon,k}(\theta)}{e^{\theta^T \cdot t(g)} }}.
\end{align*}
			Now, showing that $\lim\limits_{n \rightarrow \infty} \mathbb P_{\theta}(G \in M_{\epsilon,k}(\theta)) < 1$ is equivalent to showing  $\lim\limits_{n \rightarrow \infty} r_n > 0$. 
			
Let $N_m  = |M_{\epsilon,k}(\theta)|$ and let $U_{n,k}(\theta)= \underset{g \in \mathcal{G}_{n,k}}{\max} \theta^T \cdot t(g)$, and $L_{n,k} = \underset{g \in \mathcal{G}_{n,k}}{\min} \theta^T \cdot t(g)$. Without loss of generality we can assume that $L_{n,k}(\theta)$ is $0$. This follows from observing that $\mathbb P_{\theta}(G=g)$ is invariant under the translations of $\theta^T \cdot t(g)$ by $-L_{n,k}(\theta)$. Also, note that for any $g \in M_{\epsilon,k}(\theta)$, and any $0 < \epsilon < 1$, we have $\theta^T \cdot t(g) \leq U_{n,k}(\theta)$. Thus, we have, 
		\begin{align*}
		r_n &= \frac{\sum_{g \in \mathcal{G}_{n,k} \setminus M_k(\theta)} e^{\theta^T \cdot t(g)}}{\sum_{g \in M_{\epsilon,k}(\theta)}{\exp(\theta^T \cdot t(g)) }} \\ 
		&>\frac{\sum_{g \in \mathcal{G}_{n,k} \setminus M_k(\theta)} e^{\theta^T \cdot t(g)}}{N_m e^{U_{n,k}(\theta)}} \\
		& \geq \frac{\sum_{g \in \mathcal{G}_{n,k} \setminus M_k(\theta)}e^{L_{n,k}(\theta)}}{N_m e^{U_{n,k}(\theta)}} 
		= \frac{\sum_{g \in \mathcal{G}_{n,k} \setminus M_k(\theta)}e^{0}}{N_m e^{U_{n,k}(\theta)}} 
		=\frac{S_k(n) - N_m}{N_m e^{U_{n,k}(\theta)}} = \frac{\frac{S_k(n)}{N_m} - 1 }{e^{U_{n,k}(\theta)}}
		\geq \frac{\frac{S_k(n)}{2N_m} }{e^{U_{n,k}(\theta)}} \\
		&\geq \frac{c_0 \sqrt{S_k(n)}}{2e^{U_{n,k}(\theta)}} (\mbox{ By assumption, } N_m < c_0 \cdot \sqrt{S_k(n)})\\
		&\geq \frac{c_0}{2} \frac{\sqrt{e^{c_1 \cdot n \log n}}}{e^{U_{n,k}(\theta)}} (\mbox{ Since } \log S_k(n) > c_1\cdot n\log n, \mbox{ from Theorem \ref{thm:lowerboundSupport}}). 
		\end{align*}

The last inequality follows from Theorem \ref{thm:lowerboundSupport}, which states that there exists a constant $c_1$, and an $n_0$ such that for all $n > n_0$, $\log S_k(n) \geq c_1 \cdot n \log n$. Recall that  $t(g)$ being stable means that  for all $c> 0$, there exists an $n_0$ such that for all $n > n_0$, $\underset{g \in \mathcal{G}_{n,k}}{\max} t(g) < c \cdot \log S_k(n)$.Thus, for all $c>0$,
\begin{align*}
U_{n,k}(\theta) &= \underset{g \in \mathcal{G}_{n,k}}{\max} \theta^T \cdot t(g) \\
&< c_{\theta} \cdot c \cdot  \log ( S_k(n)) \\
&< c_{\theta} \cdot c \cdot c_2 \cdot n \log n.
\end{align*}
The last inequality again follows from Theorem \ref{thm:lowerboundSupport} which states that there exists a constant $c_2$ and an $n_0$ such that for all $n > n_0$, $\log S_k(n) \leq c_2 \cdot n \log n$. Here, $c_{\theta}$ is a constant that depends on $\theta$.	Thus we get, for all $c>0$, there exists an $n_0$, $c_1$ and $c_2$ such that for all $n > n_0$,	
\begin{align*}
		r_n &> \frac{c_0}{2}\frac{e^{\frac{c_1}{2} \cdot n \log n}}{e^{U_{n,k}(\theta)}} \\
		&>\frac{c_0}{2} \frac{e^{\frac{c_1}{2} \cdot n \log n}}{e^{c \cdot c_{\theta} c_2 \cdot n \log n}}\\
		&>\frac{c_0}{2} e^{\left(\frac{c_1}{2} - c \cdot c_2 c_{\theta}\right) \cdot n \log n}.
\end{align*} 
Since this holds for any $c>0$, let us choose 
%\sout{\color{magenta}{$c$ such that $\frac{c_1}{2} - c\cdot c_2 c_{\theta} \geq 0$, or $c \leq \frac{c_1}{2 c_2 c_{\theta}}$ which implies $\lim r_n \rightarrow \infty$.}}
$c$ such that $\frac{c_1}{2} - c\cdot c_2 c_{\theta} = 0$. Then,  $r_n > \frac{c_0}{2}>0$ in the limit, as required. 
\end{proof}

In order to show an explicit example of a model for which we can find a set of parameter values $\theta$ for which  Theorem~\ref{thm:non-degenerate} holds, we spell out the result for the example of the triangle DERGM studied in the previous section. At the same time we can prove stronger result, relaxing the assumption on the degeneracy $k$. 
\begin{corollary}[Stability implies non-(near)-degeneracy for edge-triangle DERGM]
	\label{thm:non-degeneratev2}
	Consider DERGM with parameter vector $\theta= (\theta_1, \theta_2)$ and sufficient statistics $(e(g),\triangle(g))$. 
 	Allow the degeneracy parameter $k$ to increase as follows:
	\begin{enumerate}
		\item $k = o(\sqrt{\log n})$\label{kGrowth}. 
	\end{enumerate} 
		For $\theta\in\Theta$, suppose that:
	\begin{enumerate}
		\item $|\theta|_1 < o(\log n)$, where $|\theta|_1$ is the $l_1$ norm of $\theta$,
		\item $\theta\in\Theta$ is such that there exists and constant $c_0$ and an $n_0$ such that for all $n> n_0$, $|M_{\epsilon,k}(\theta)| < c_0 \sqrt{S_k(n)}$, that is the number of graphs in the set of $\epsilon$ modes does not grow larger than the square root of the total number of graphs in the support of the DERGM.\label{modesGrowth}
	\end{enumerate} 

	Then, the edge-triangle DERGM is asymptotically non-(near)-degenerate at $\theta$. 
\end{corollary}

Assumption \ref{kGrowth} of course holds for fixed values of $k$, thus it is not restrictive on the DERGM as we defined it, but rather is a relaxation. 
The last assumption %\ref{modesGrowth}
  is the same as in the theorem above. 
Note that   the former  (concerning the growth of $k$) is weak, whereas the latter  (concerning the number of modes) is strong.

\begin{proof}
To prove asymptotic non-(near-)degeneracy, we repeat the same steps as in the theorem above, but consider a finer lower bound on the ratio $r_n$ from the end of the proof: 		
				\begin{align*}
		r_n > \frac{c_0}{2} \cdot \frac{e^{\frac{c_1}{2} \cdot n \log n}}{e^{U_{n,k}(\theta)}}. 
		\end{align*} 

		Now, let us examine $r_n$ for the case of number of edges and triangles. From Proposition \ref{prop:stableSufficientStatistics}, there exists a constant $c_2$ and an $n_0$ such that for all $n> n_0$, the following holds: 
		\begin{align*}
		U_{n,k}(\theta) &= \max_g (\theta_1, \theta_2)^T \cdot (e(g),\triangle(g)) < |\theta|_1 \cdot \max_g (e(g)+\triangle(g)) \\
		&< |\theta|_1 \cdot c_2 \cdot k^2 \cdot n 
		\end{align*}
		Thus we have,
		\begin{align*}
		r_n &\geq \frac{c_0}{2} \cdot \frac{e^{\frac{c_1}{2} \cdot n \log n}}{e^{U_{n,k}(\theta)}} \\
		&\geq \frac{c_0}{2} \cdot \frac{e^{\frac{c_1}{2} \cdot n \log n}}{e^{|\theta|_1 \cdot c_2\cdot k^2 \cdot n}}.
		\end{align*}
		If we allow $|\theta|_1 =o( \log n)$,  and $k = o(\sqrt{\log n})$, then we have $c_2 |\theta|_1 \cdot k^2 \cdot n = o(n \log n)$, which means for all $c>0$, there exists an $n_0$ such that for all $n> n_0$, $c_2 |\theta|_1 \cdot k^2 \cdot n < c \cdot n \log n$. Thus, we have,
		\begin{align*}
		r_n 
		&\geq \frac{c_0}{2} \cdot \frac{e^{\frac{c_1}{2} \cdot n \log n}}{e^{c \cdot n \log n }}.
		\end{align*}
Choosing $c = \frac{c_1}{2}$, we get $r_n \geq \frac{c_0}{2}$, as needed.

\end{proof}
Corollary \ref{thm:non-degeneratev2} shows that the edge-triangle DERGM is asymptotically non-(near)-degenerate for $k = o(\log n)$ and $|\theta|_1 = o(\log n)$.
This result implies that for large $n$, the edge-triangle DERGM cannot place all its mass on the set of $\epsilon$-modes, and there must be a considerable amount of mass assigned to points outside the set of $\epsilon$-modes.

\section{Maximum Likelihood Estimation of DERGMs}\label{sec:MLEofDERGM}

In this section, we consider the problem of estimating the parameters of a DERGM given by Equation~\eqref{eq:DERGM} from a single observed graph $g_{obs}$  on $n$ nodes. Suppose that $g_{obs}$ has degeneracy $k_{obs}$. 
 To fit a DERGM to $g_{obs}$, we need to estimate the parameter vector $\theta$ and the degeneracy parameter $k$. 
 From now on, we assume $k$ is fixed and equal to $k_{obs}$; see Remark~\ref{rmk:whyKobserved}. For a fixed $k$, one can write the log-likelihood function of a DERGM in the following form:
\begin{align}
\label{eq:likelihood}
l_k(\theta;g_{obs}) = -\log \left(\underset{g \in \mathcal{G}_{n,k}}{\sum} {\exp\left(\theta^T\Delta(g;g_{obs}) \right)}\right), 
\end{align}
where $\Delta(g;g_{obs}) = t(g) - t(g_{obs})$. We will also use $\Delta(g)$ to denote $\Delta(g;g_{obs})$ when it is clear that $g_{obs}$ is fixed. The maximum likelihood estimate of $\theta$ is 
$$
\hat \theta = \arg\max l_k(\theta; g_{obs}).
$$
As is the case with ERGMs, directly maximizing Equation~\eqref{eq:likelihood} to obtain $\hat \theta$ is intractable. Hence, we need to resort to \emph{approximate} maximization. The most commonly used method is the MCMC-MLE proposed in \cite{geyer1992constrained} and applied to ERGMs by and \cite{HunterandHandcock}. An alternative is to use stochastic approximation of \cite{robbins1985stochastic}, see \cite{snijders2002markov}. However, as stated in \cite{hunter2012computational}, and shown in \cite{geyer1992constrained}, the MCMC-MLE procedure makes more efficient use of the samples in comparison to the stochastic approximation method.  

Therefore, to estimate DERGMs, we use the MCMC-MLE method, combined with the step length algorithm of \cite{hummel2012improving}. The key idea in MCMC-MLE is to approximate the log-likelihood function using importance sampling, which is then maximized to obtain an approximate MLE. The approximate MLE is used to sample graphs and obtain an improved approximation of the likelihood function, which is again maximized. This process is repeated iteratively, until convergence.

More specifically, letting $\theta_0$ be a fixed starting value (usually taken to be the maximum pseudo-likelihood estimator), the log-likelihood from Equation~\eqref{eq:likelihood} can be written as:  
\begin{align}
\label{eq:Expectedloglikelihood}
l_k(\theta;g_{obs}) = -\log \left(c_k(\theta_0)\right) - \log \mathbb E_{\mathbb P_{\theta_0,k}} \left[ \exp((\theta-\theta_0)^t \Delta(G;g_{obs}))\right], 
\end{align}
where $\Delta(G; g_{obs}) = t(G) - t(g_{obs})$ and the expectation is over $\mathbb P_{\theta_0,k}$, which denotes a DERGM with parameters $\theta_0$ and degeneracy parameter $k$.
If $G_1, \ldots, G_B$ are iid samples from $\mathbb P_{\theta_0,k}$, one can obtain a strongly consistent estimate of the log-likelihood by using 
\begin{align}
\label{eq:estloglikelihood}
\hat l_k(\theta;g_{obs}) 
&= -\log \left(c_k(\theta_0)\right) - \log \sum_{b=1}^B \left[ \exp((\theta-\theta_0)^t \Delta(G_b;g_{obs}))\right] + \log B \\ \nonumber
&\propto  \log \sum_{b=1}^B \left[ \exp((\theta-\theta_0)^t \Delta(G_b;g_{obs}))\right].
\end{align}
The estimated log-likelihood in Equation~\eqref{eq:estloglikelihood} is maximized to obtain an approximate maximum likelihood estimator. Thus, the approximate MLE is defined as
\begin{equation}\label{eq:thetaTilde}
\tilde{\theta} = \arg\max \hat l_k(\theta, g_{obs}).
\end{equation}

In general, it is not possible to obtain iid samples from $\mathbb P_{\theta_0}$, and one resorts to MCMC methods to draw approximate samples from the model by running the Markov chain until convergence, see \cite{snijders2002markov} and \cite{HunterandHandcock} for more details. Thus, the key step in estimating DERGMs using MCMC-MLE is to draw MCMC samples from a DERGM with a fixed value of $\theta$ with the support restricted to $k$-degenerate graphs.

\subsection{Sampling graphs from a DERGM with a fixed parameter}
\label{sec:modelSampler}

 In this section, we discuss an MCMC algorithm for sampling graphs from the DERGM for a fixed value of $\theta$ with degeneracy parameter $k$. The key issue is that to sample from a DERGM using MCMC, we need to ensure that the proposed graphs are in the set $\mathcal G_{n,k}$, i.e. they have degeneracy restricted to $k$.  To this end, we consider two different approaches: the first, straightforward approach, is to use the usual tie-no-tie proposal (see, for example, \cite{caimo2011bayesian}) along  with the Metropolis-Hastings step.  Such a proposal may generate graphs outside the set $\mathcal G_{n,k}$, which are naturally rejected by the Metropolis-Hastings algorithm. Thus, whenever the degeneracy of the proposed graph is more than $k$, the graph is rejected, otherwise it is accepted with the usual acceptance probability that depends on the change statistics, see \cite{hunter2008ergm} for more details. Note that the degeneracy of a graph can be computed in $O(m)$ time, where $m$ is the number of edges, using the algorithm of \cite{batagelj2003m}. 
 
 While the first method works, it can be wasteful and slow, i.e. at each step of the Markov chain, we have to compute the degeneracy of the graph and reject it whenever it is larger than $k$. The second approach is to directly propose graphs from the set $\mathcal G_{n,k}$. For this, we develop a uniform sampler that proposes graphs uniformly from the set of all $k$-degenerate graphs. The uniform sampler is presented in section \ref{sec:uniformSampler}.

Algorithm \ref{alg:sampleModel} summarizes the approach 2  where the proposal is the uniform distribution from $\mathcal G_{n,k}$, denoted by $\mathcal U_{n,k}$. Let $\pi(g) \propto \exp(\theta_{0}^t t(g))$. The Metropolis-Hastings acceptance ratio becomes
$$\alpha(g_{current}, g_{proposed}) = \min \left(1, \frac{\pi(g_{proposed})}{\pi(g_{current})} \right).$$

{\footnotesize 
	\begin{algorithm}[H]
		\label{alg:sampleModel}
		\LinesNumbered
		\DontPrintSemicolon
		\SetAlgoLined
		\SetKwInOut{Input}{input}
		\SetKwInOut{Output}{output}
		\Input{$g_0$, the starting value of the chain}
		%	\Output{}
		\BlankLine
		Let $g_0$ be the starting value of the chain and set $g_{current} = g_0$.\;
		For $t=1,\ldots, B$:\;
		Propose a new value $g_{proposed}$ from $\mathcal U_{n,k}$\;
		Define $$\alpha(g_{current}, g_{proposed}) = \min \left(1, \frac{\pi(g_{proposed})}{\pi(g_{current})} \right).$$ \;
		Let $u \sim Unif(0,1)$.\;
		If $ u \leq \alpha$, accept the new proposal and set $g_{t+1} = g_{proposed};$\;
		Else set $g_t = g_{current}.$\;
		\caption{Independent Metropolis algorithm to sample from the model} 
	\end{algorithm}
}

\subsection{Existence of MLE and the approximate MLE} 
\label{sec:existence}
There are two likelihood functions: the true likelihood $l(\theta)$ given by Equation \eqref{eq:likelihood} and the estimated likelihood $\hat l(\theta)$  given by Equation \eqref{eq:estloglikelihood}. Correspondingly, there are two maximizers, the true MLE $\hat \theta$ and the approximate MLE $\tilde \theta$. We will discuss the existence of the true MLE and the approximate MLE and argue that using a smaller $k$ makes the estimation of the MLE easier. 

Using the standard theory of exponential families \cite{barndorff2014information}, existence of the true MLE $\hat \theta$ depends on the marginal polytope, that is,  the convex hull of sufficient statistics of the set $\Gnk$. %$G_{n,k}$. 
 The log-likelihood function is concave and a unique maximum exists if and only if the observed sufficient statistic $t(g_{obs})$ lies in the relative interior of the marginal polytope. The marginal polytopes of ERGMs are difficult to obtain in general (see for example \cite{engstrom2010polytopes}) and known only in few special cases, such as  \cite{rinaldo2013maximum}, \cite{karwa2016inference}. Obtaining the marginal polytopes for the degeneracy-restricted ERGMs appears to be more difficult and is an open problem in general, as it can only be computed for one specific DERGM at a time. We will compute these polytopes numerically for the  edge-triangle DERGM in Section~\ref{sec:simulationsOnModelBehavior}.

On the other hand, existence of the approximate MLE can be checked numerically. As discussed in \cite{Handcock03assessingdegeneracy}, the estimated log-likelihood \eqref{eq:estloglikelihood} can be written as the log-likelihood of a model from a discrete exponential family with support over $t(G_1), \ldots, t(G_B)$ with observed sufficient statistic $t(g_{obs})$. Hence, using again the standard theory of exponential families \cite{barndorff2014information}, one can show that the estimated log-likelihood is concave and Equation~\eqref{eq:estloglikelihood} has a unique maximum if and only if $0$ lies in the interior of the convex hull of $\{\Delta(G_1,g_{obs}), \ldots, \Delta(G_B,g_{obs})\}$. 
Thus, assuming that the MLE exists, the existence of the approximate MLE is crucially tied to the sampling algorithm used to approximate the likelihood, which in turn depends on the behavior of the model. 

%%%%%%=====================%%%%%%
\section{Simulations on the effect of $k$ on model behavior}%The choice of $k$ and model behavior: simulations for the edge-triangle DERGM}
\label{sec:simulationsOnModelBehavior}
\label{sec:simulations}

In this section, we use extensive simulations to show that ``bad behavior'' of the model is a function of the degeneracy parameter.  In particular, the bad behavior of the model increases with values of degeneracy parameter $k$, where ``bad behavior`` is an umbrella term used to denote model degeneracy, sensitivity, the difficulty of MLE computations. These simulations provide additional justification to the theory developed in Section \ref{sec:stability} and illustrate that restricting the support of the model to $k$-degenerate graphs improves model behavior. We focus on the edge-triangle DERGM as a running example, a model whose sufficient statistics are the number of edges and the number of triangles of the graph. To illustrate the changing behavior of the degeneracy-restricted ERGMs, in each of the following examples we fix $n$ and vary $k$ from the observed value to the maximum $k=n-1$. 

\begin{remark}\rm 
The edge-triangle model is also the running example in  \cite{RinaldoFienbergZhu09}, where the authors show that the model degeneracy is captured by polyhedral geometry of the model and the entropy function. We also study the model polytope and the entropy function of DERGMs. 
\end{remark}

\subsection{Insensitivity and lack of degeneracy of DERGMs}

We begin by studying the effect of $k$ on the mean value and the natural parameters of DERGMs. The goal is to gain insight into the \emph{model degeneracy} and \emph{excessive sensitivity} of DERGMs as a function of $k$. Roughly, the model is said to suffer from \emph{degeneracy} issues, if the mean value parameters of the model are pushed to the boundary for different values of the natural parameter. Similarly, the model is said to suffer from \emph{excessive sensitivity}, small changes in the values of the natural parameters lead to large changes in the mean value parameter, see \cite{schweinberger2011instability} for more details.

\begin{remark} We want to note that the term ``degeneracy'' is being used in two different contexts. In section \ref{sec:stability}, we defined \emph{asymptotic degeneracy} to denote the situation where a distribution places most of its mass on its modes. In this section, the term ``degeneracy'' is used to denote the situation when the mean value parameter of a distribution is pushed to its boundary. In fact, the second type of degeneracy is implied by asymptotic degeneracy, as shown in \cite{schweinberger2011instability}.
\end{remark}

In the rest of the section, we focus on one-parameter exponential families. We will work with normalized sufficient statistics.  Specifically, let $U_k$ denote the maximum of $t(g)$ when $g \in G_{n,k}$. Let the normalized sufficient statistic be $t_{norm}(g) = t(g)/U_k$.
For the natural parameter $\theta$,  the \emph{mean value parameter} is given by $ \mu_k(\theta) = \mathbb E_{\mathbb P_{\theta, k}} t_{norm}(g)$. %, where $t_{norm}(g)$ is the corresponding normalized sufficient statistic.

We consider two different DERGM models: the two-star DERGM with the number of two-stars as the sufficient statistic, and the triangle DERGM with the number of triangles as the sufficient statistic. 
 Degeneracy corresponds to the situation where if $\theta > 0$, $\mu_k(\theta) \rightarrow 1$ and $\theta <0$, $\mu_k(\theta) \rightarrow 0$. Sensitivity corresponds to the situation where the derivative of $\mu_k(\theta)$ with respect to $\theta$ is very large in a small neighborhood of $\theta$.

\begin{remark}
	When $k=n-1$, from the properties of standard exponential families, we can show that the derivative of $\mu_k(\theta)$ with respect to $\theta$ is the variance of the sufficient statistic. 
	 Thus, another way to view sensitivity is that the variance of the sufficient statistic is very large in a small neighborhood of $\theta$. \cite{fellows2017removing} restrict the variance, addressing the degeneracy and sensitivity issues.
\end{remark}
Recall that our goal is to study the map from $\theta$ to $\mu_k(\theta)$ for varying values of $k$ and gain insights into model behavior. To avoid any issues due to MCMC sampling, we compute this map exactly for a small network, where enumeration is possible.  Specifically, we consider networks defined on $n=7$ nodes. When $n=7$, there are a total of $2^{7 \choose 2}$ possible simple networks. We enumerate all possible networks, and compute the number of edges, two-stars, triangles and degeneracy of each network. The total number of networks with different degeneracy values is shown in Table \ref{tab:smallndgen}. 

\begin{table}[ht]
	\centering
	\begin{tabular}{r@{\hskip 0.1in}r@{\hskip 0.1in}r@{\hskip 0.1in}r@{\hskip 0.1in}r@{\hskip 0.1in}r@{\hskip 0.1in}r}
		\hline
		$k$& 1 & 2 & 3 & 4 & 5 & 6 \\ 
		\hline
		 $n(g)$ & 36960 & 1095461 & 900298 & 63801 & 630 &   1 \\ 
		\hline
	\end{tabular}
	\caption{Number of graphs of degeneracy exactly $k$ for $n=7$ nodes}
\label{tab:smallndgen}
\end{table}

\begin{figure}
	\begin{subfigure}[b]{0.25\textwidth}
		\includegraphics[width=\textwidth]{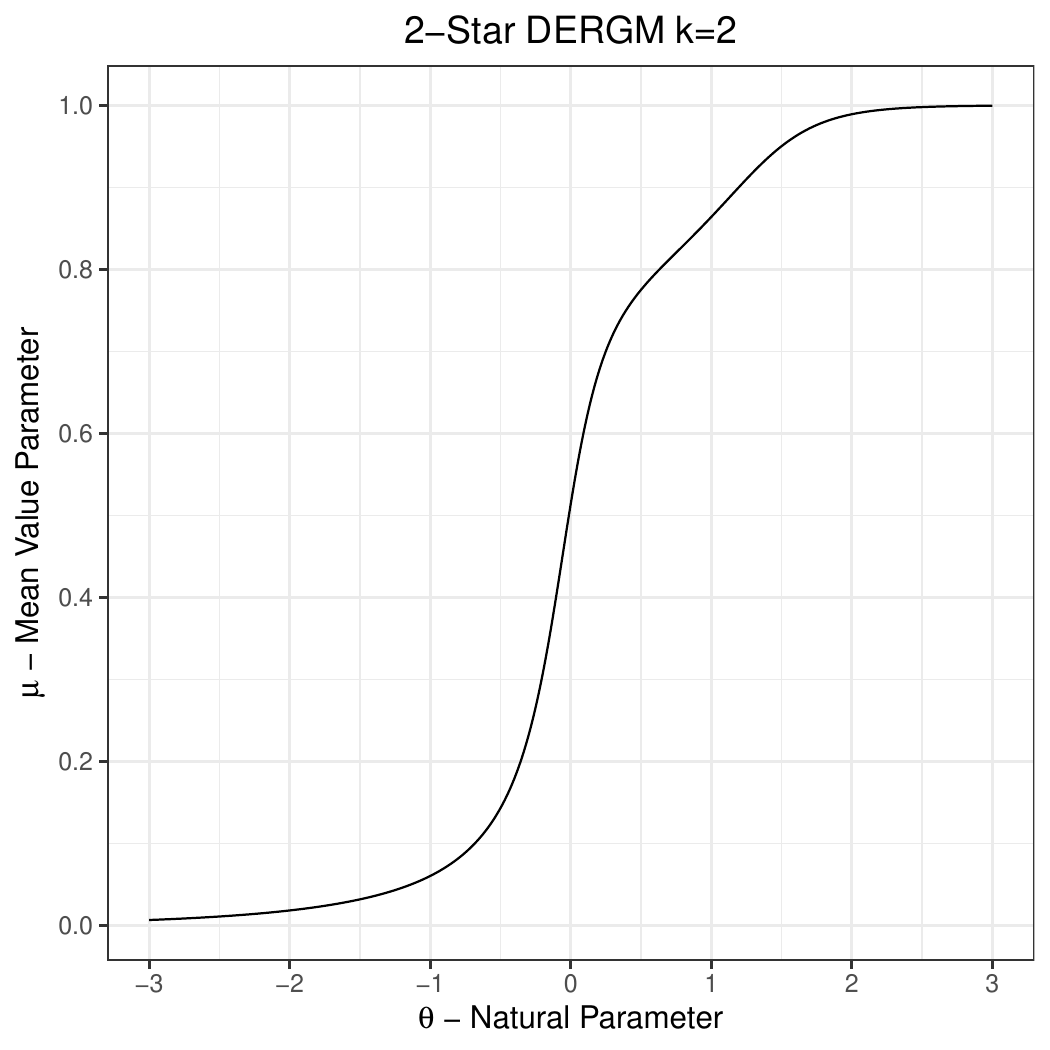}
		\caption{$n=7$, $k=2$}
		\label{fig:1star}
	\end{subfigure}
	\begin{subfigure}[b]{0.25\textwidth}
		\includegraphics[width=\textwidth]{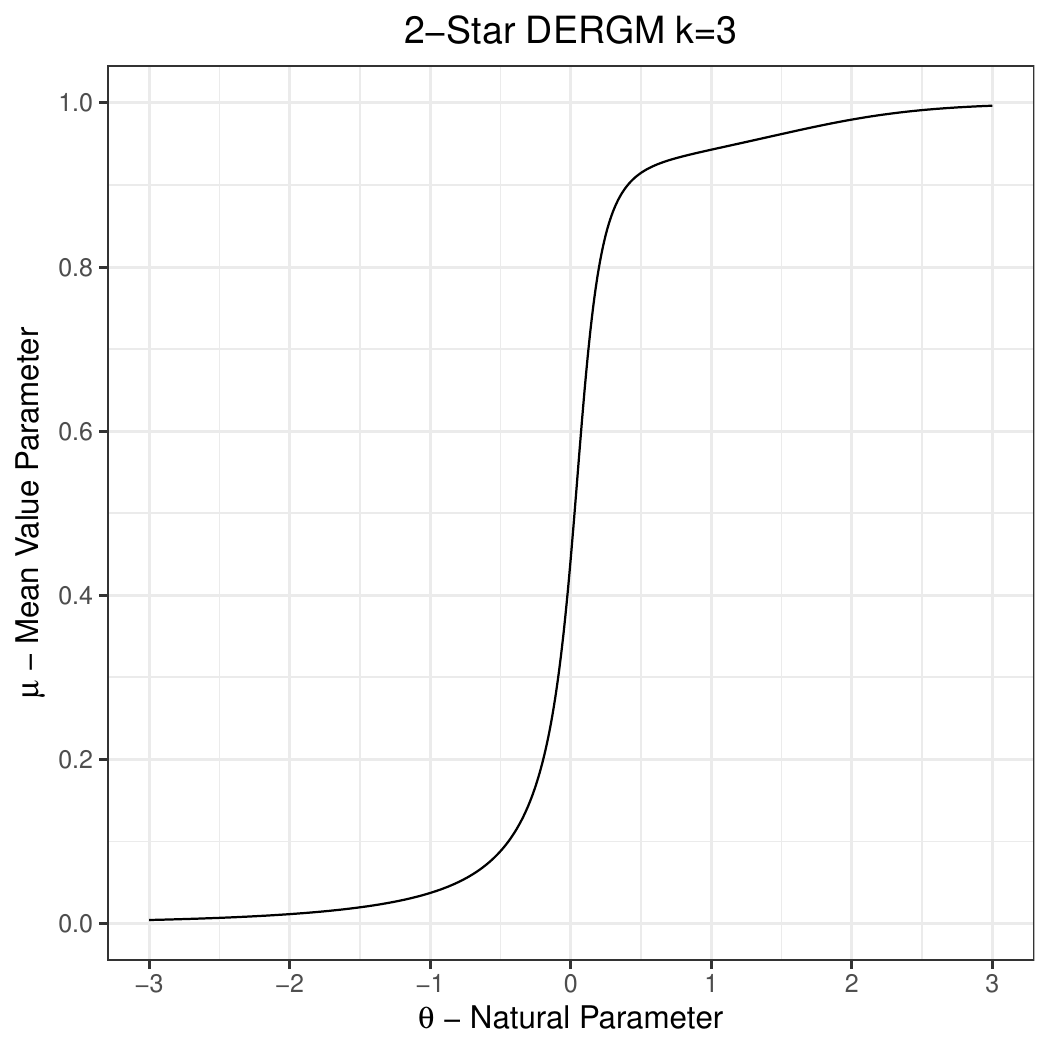}
		\caption{$n=7$, $k=3$}
		\label{fig:2star}
	\end{subfigure}
	\begin{subfigure}[b]{0.25\textwidth}
		\includegraphics[width=\textwidth]{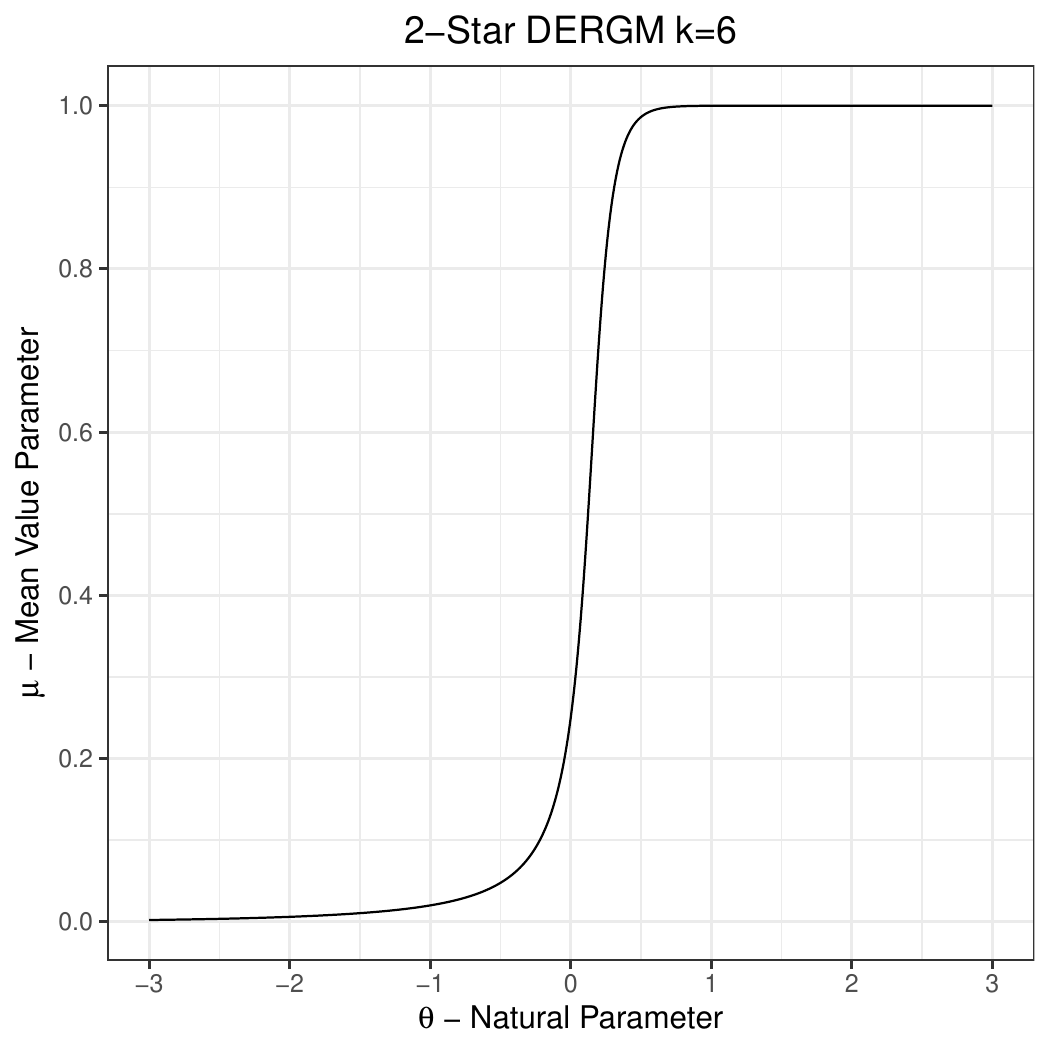}
		\caption{$n=7$, $k=6$}
		\label{fig:5star}
	\end{subfigure}
\caption{Mean value Parameters vs Natural parameters for the 2-star DERGM for $n=7$ and $k=2,3,6$ respectively.}
	\label{fig:meanvalue2stars}
\end{figure}

\begin{figure}
  \begin{subfigure}[b]{0.25\textwidth}
    \includegraphics[width=\textwidth]{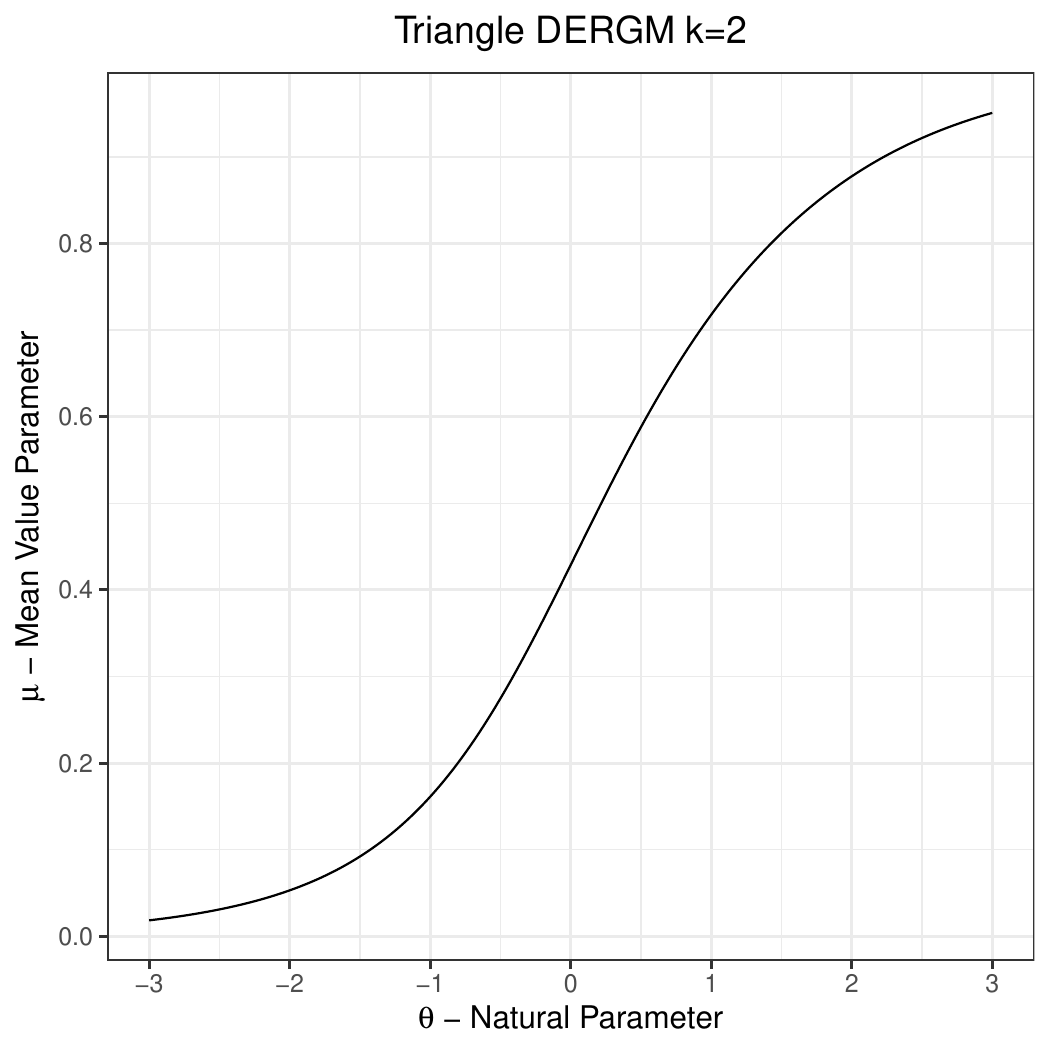}
    \caption{$n=7$, $k=2$}
    \label{fig:1tri}
  \end{subfigure}
  \begin{subfigure}[b]{0.25\textwidth}
    \includegraphics[width=\textwidth]{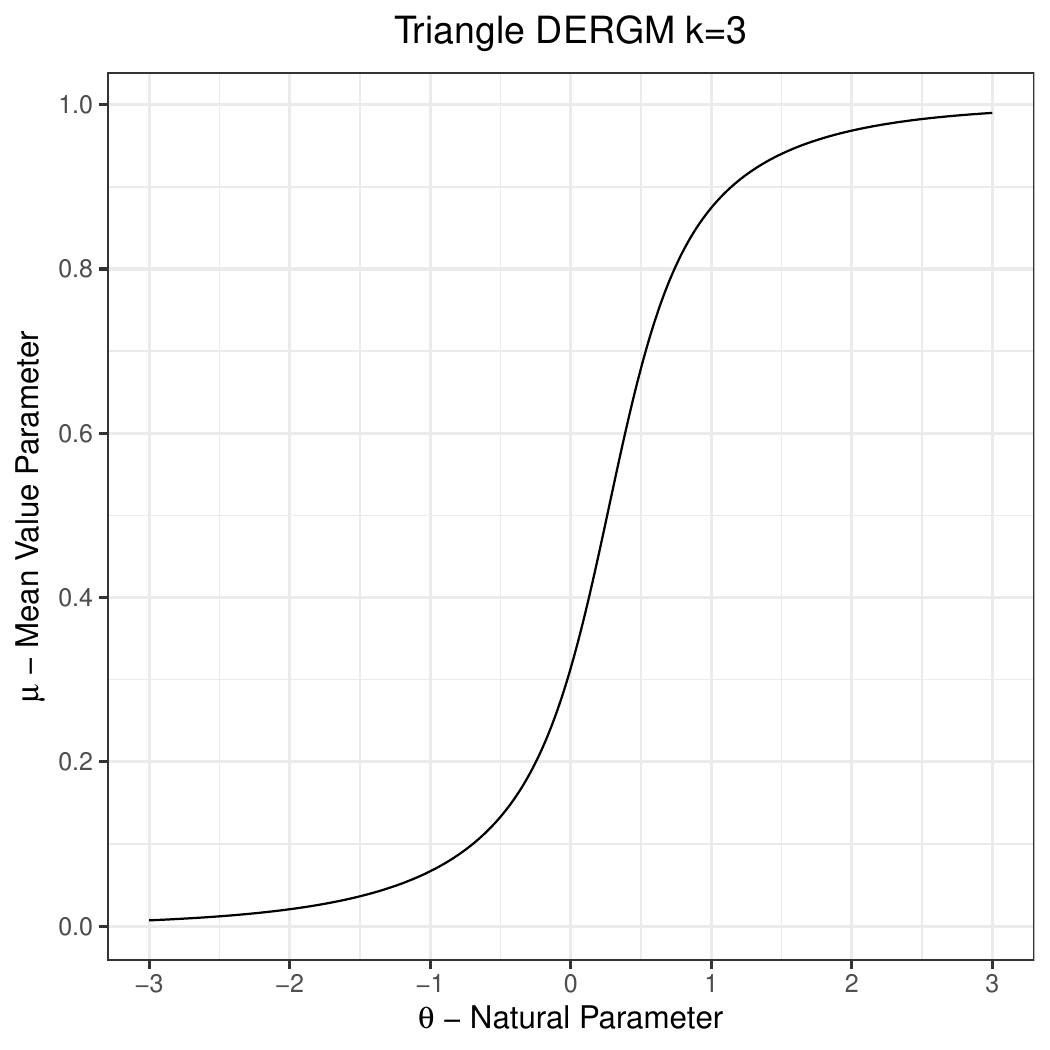}
    \caption{$n=7, k=3$}
    \label{fig:2tri}
  \end{subfigure}
\begin{subfigure}[b]{0.25\textwidth}
	\includegraphics[width=\textwidth]{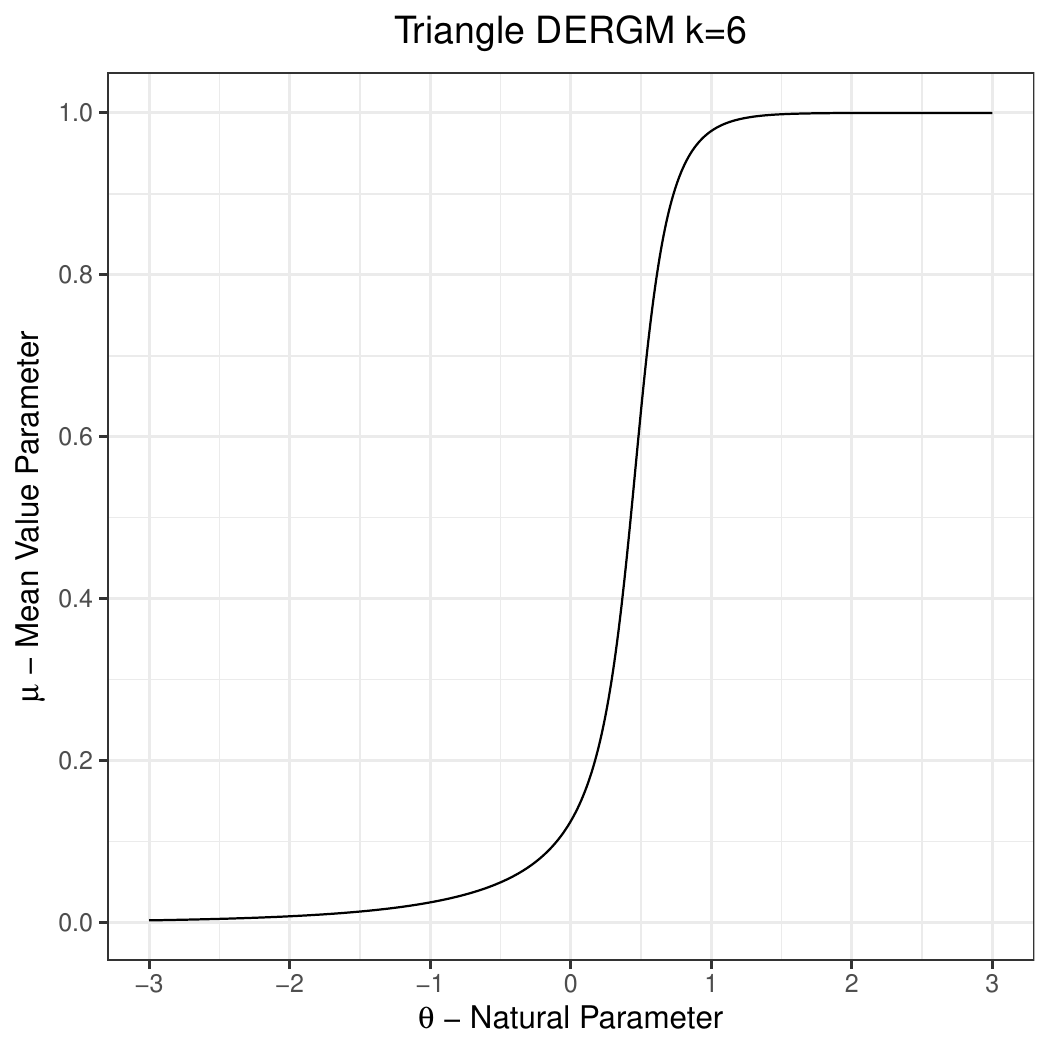}
	\caption{$n=7, k=6$}
	\label{fig:3tri}
\end{subfigure}
\caption{Mean value parameters vs Natural parameters for the triangle DERGM for $n=7$ and $k=2,3,6$ respectively.}
	\label{fig:meanvalueTriangles}
\end{figure}

The plot of mean value vs natural parameter for each DERGM model is generated as follows. We fix a value of $k$, and fix a sufficient statistic. Next, we vary $\theta$ from $-3$ to $3$ in steps of $0.01$. For each value of $\theta$, we compute the corresponding mean value parameter $\mu_k(\theta)$ using the enumerated networks.  We normalize $\mu_k(\theta)$ to make sure it lies between $0$ and $1$ and plot the normalized $\mu_k(\theta)$ on $y$-axis and the natural parameter $\theta$ on the $x$-axis. We repeat this process for different values of $k$, and obtain a separate plot for each value of $k$. Similarly, we get different sets of plots for each DERGM. The results are shown in Figures \ref{fig:meanvalue2stars} and \ref{fig:meanvalueTriangles}.

\begin{figure}
	\begin{subfigure}[b]{0.4\textwidth}
		\includegraphics[width=\textwidth]{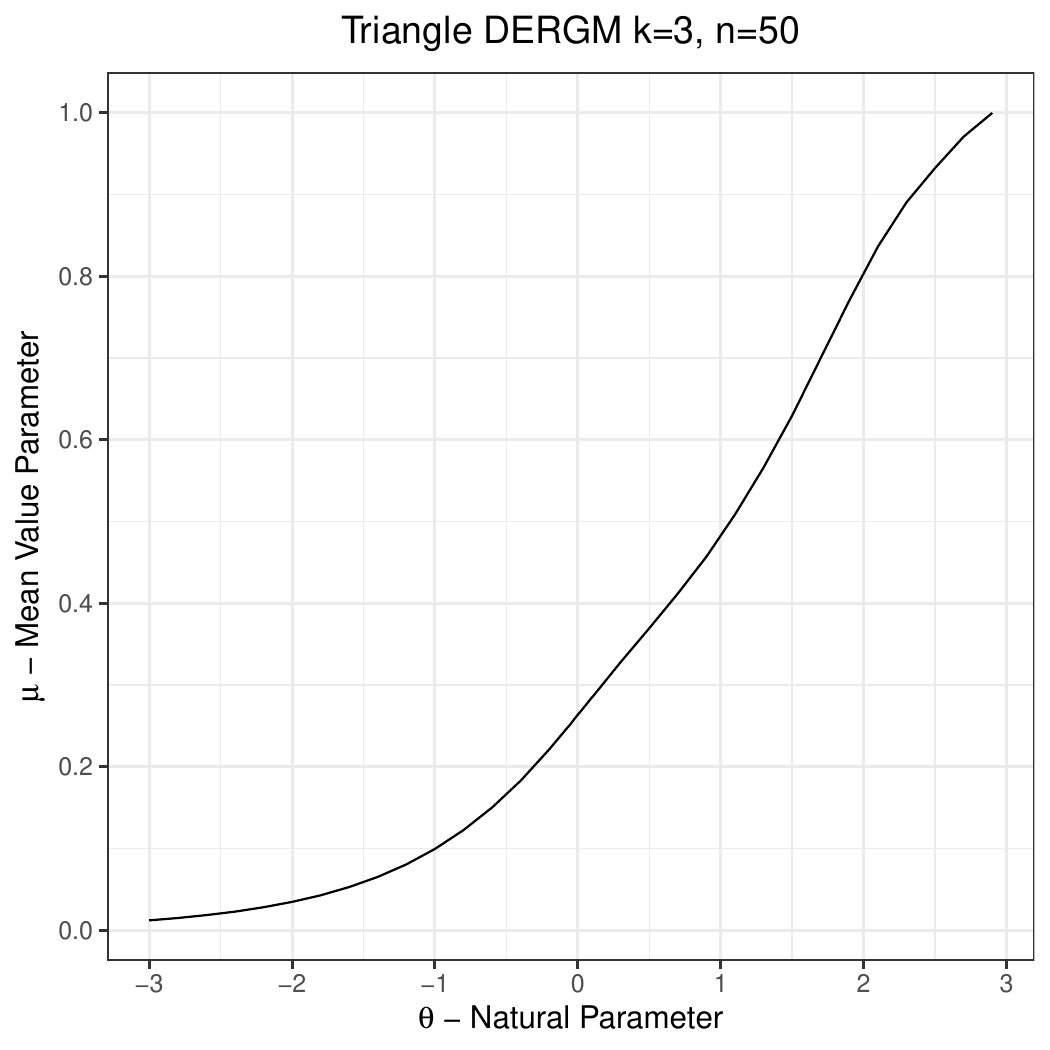}
		\caption{$n=50$, $k=3$}
		\label{fig:1tri50}
	\end{subfigure}
	\begin{subfigure}[b]{0.4\textwidth}
		\includegraphics[width=\textwidth]{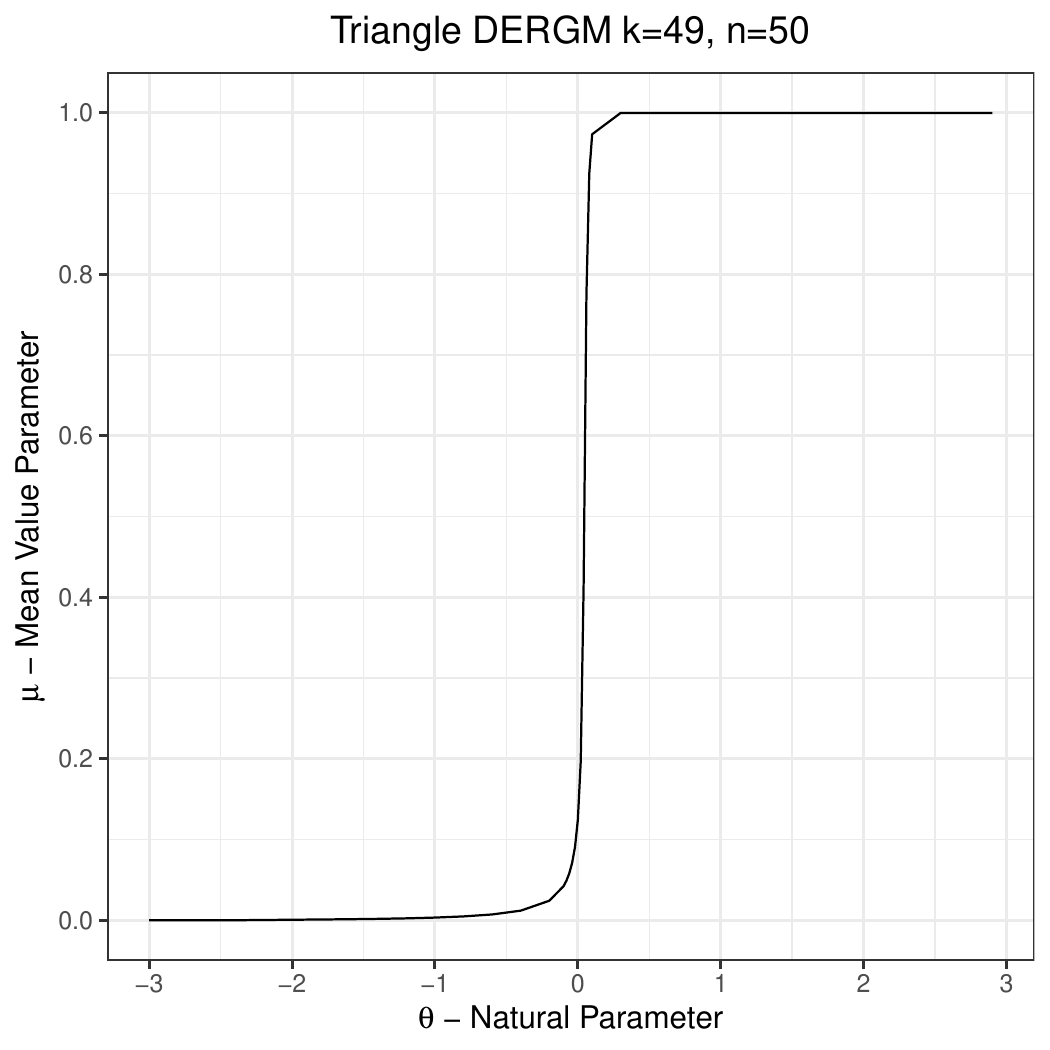}
		\caption{$n=50, k=49$}
		\label{fig:2tri50}
	\end{subfigure}
	\caption{Simulated plot of Mean value parameters vs Natural parameters, based on MCMC, for the triangle DERGM for $n=50$ and $k=3$ and $k=49$ respectively.}
	\label{fig:meanvalueTrianglesSimulated}
\end{figure}

Let us focus on Figure \ref{fig:3tri}. This figure shows the map between $\theta$ and $\mu_k(\theta)$ for the triangle-DERGM when $k=6$ and $n=7$, which is the same as the ERGM (since $k=6$ is the maximum possible, there is no support restriction). The plot shows that the mean value parameter is pushed to its corresponding boundaries for positive and negative values of $\theta$, i.e. for $\theta > 0$, $\mu_k(\theta)$ is close to 1, and for $\theta < 0$, $\mu_k(\theta)$ is close to 0. Moreover, for $\theta$ close to $0$, the mean value parameter is very sensitive to small changes in $\theta$. This is the classic model degeneracy and excessive sensitivity. On the other hand, if we consider Figures \ref{fig:1tri} and \ref{fig:2tri}, we can see that if we restrict the support to $2$-degenerate graphs or $3$-degenerate graphs, the mean value map improves. Specifically, for $k=2$, Figure \ref{fig:1tri} shows that $\mu_k(\theta)$ is not pushed to its boundaries for positive or negative values of $\theta$, and has a small derivative near $\theta=0$. This shows that the model does not suffer from degeneracy and excessive sensitivity when $k$ is small.  A similar conclusion holds for the $2$-star model shown in Figure \ref{fig:meanvalue2stars}. %Recall that for any graph $g$, the maximum likelihood estimate of the mean value parameter is equal to the observed sufficient statistic. Thus, this 
We also created such plots for  $n=50$, for which we had to resort to MCMC sampling to estimate the mean value parameters, see Figure \ref{fig:meanvalueTrianglesSimulated} for the triangle DERGM for $k=3$ and $k=50$. The results for this setting was the same as described here: For small values of $k$, the triangle and the two-star DERGM does not suffer from excessive sensitivity and model degeneracy.

\subsection{Existence of  approximate MLE, the model polytope, and entropy}
\label{sec:simulationsPolytopesAndEntropy}

Consider first the issue of existence of the approximate MLE. Recall from Section~\ref{sec:existence} that in the MCMC-MLE estimation, the approximate MLE does not exist when the observed sufficient statistics lies outside of the convex hull of the %\sout{estimated model polytope} 
sufficient statistics sampled from $\mathbb P_{\theta_0}$. 
In DERGMs, this is more likely to happen when the degeneracy parameter $k$ is large relative to the observed graph degeneracy. 

\begin{figure}[H]
	\begin{subfigure}[b]{0.45\textwidth}
	\centering 
		\includegraphics[scale=0.3]{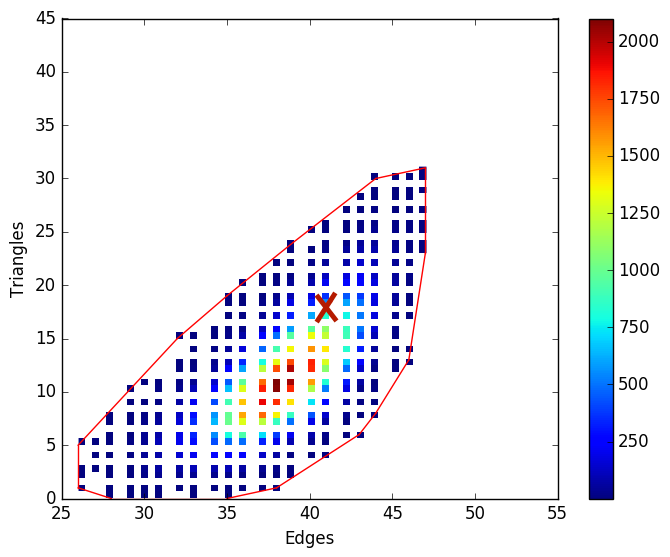}
		\subcaption{$n=18$, $k=3$}
	\end{subfigure}
	\begin{subfigure}[b]{0.45\textwidth}
	\centering 
		\includegraphics[scale=0.3]{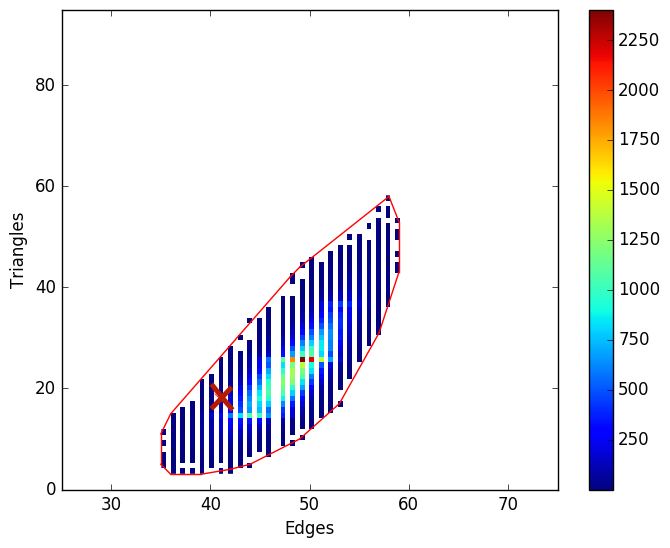}
		\subcaption{$n=18$, $k=4$}
	\end{subfigure}
	\begin{subfigure}[b]{0.45\textwidth}
	\centering 
		\includegraphics[scale=0.3]{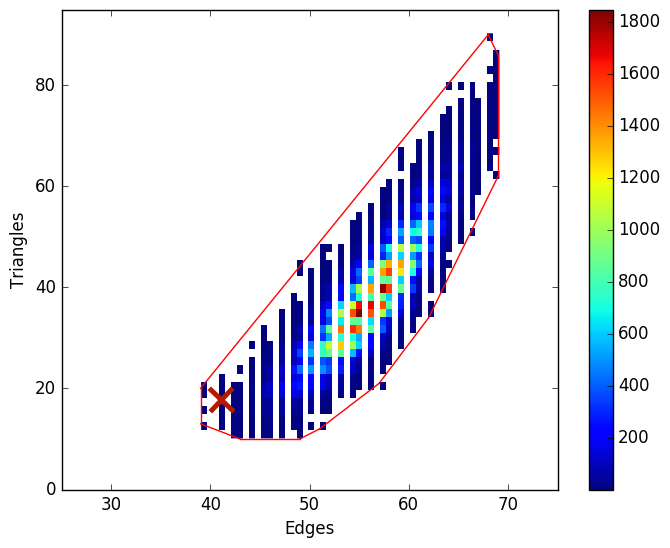}
		\subcaption{$n=18$, $k=5$}
	\end{subfigure}
	\begin{subfigure}[b]{0.45\textwidth}
	\centering 
		\includegraphics[scale=0.3]{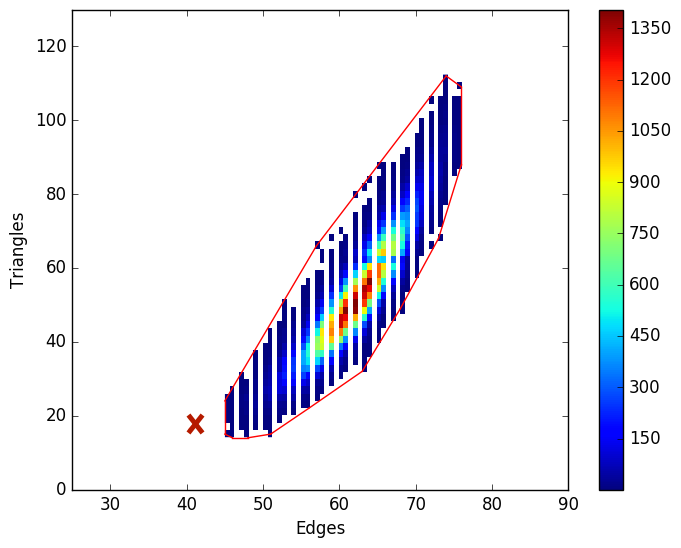}
		\subcaption{$n=18$, $k=6$}
	\end{subfigure}
		\caption{Estimated edge-triangle DERGM model polytopes for increasing $k$, with $x$ marking the location of the observed sufficient statistic of the Sampson graph. Sample size  is $100,000$ each; we have verified that the results do not change when sample size is increased to $1,000,000$. We do not plot the estimated polytopes for all other values of $k>6$, but the reader can rest assured that  the observed value of the sufficient statistics of the Sampson graph only gets farther removed from the convex hull.}
	\label{fig:polytopesWithChangingK}
\end{figure}

As an example to illustrate this phenomenon, consider fitting the edge-triangle DERGM to Sampson monastery data \cite{Sampson68}, in particular, the time period T4, available at \cite{Pajek} and \cite{hunter2008ergm}. 
In this data set, $n=18$ and observed graph degeneracy is $k=3$. 
Building on the correspondence between MLE non-existence and the model polytope from \cite{RinaldoFienbergZhu09}, we study the location of the observed edge-triangle vector with respect to estimated DERGM model polytopes for varying values of $k$. Recall that the model polytope for an exponential family model is the convex hull of all observable vectors of sufficient statistics. We estimate the DERGM polytopes as convex hulls of edge-triangle pairs of networks obtained by sampling graphs uniformly from the support $\Gnk$, using Algorithm~\ref{alg:uniformGnk}. 
Figure~\ref{fig:polytopesWithChangingK} shows the estimated model polytopes for different values of $k$, along with the relative location of the Sampson edge-triangle vector. % sufficient statistic of the Sampson network. 
  When $k=3$, the observed sufficient statistic lies well in the relative interior of the sampled sufficient statistics. On the other hand, when $k=6$ and higher,  the observed sufficient statistic lies well outside the convex hull. 
\noindent As $k$ increases, the observed edge-triangle count is progressively pushed out of the estimated polytope and becomes probabilistically less likely under the uniform distribution (blue corresponds to lower probability). This is because for larger $k$, the uniform sampler places more weight on edge-triangle counts of denser graphs, making more sparse edge-triangle counts such as those from Sampson graph probabilistically less likely to appear. 
%{\color{red}CAN WE DELETE THIS??\sout{(note that a uniform distribution on the graphs does not imply a uniform distribution on the edge-triangle counts).}}
 Thus, for larger $k$, the observed edge-triangle count of the Sampson graph lies in the tails of the distribution induced by the uniform sampler. This in turn effects the MCMC-MLE as follows: For larger $k$, the observed sufficient statistic lies close to the boundary of the true model polytope, or as the figures show, outside the estimated polytope. Unless the MCMC algorithm finds a $\theta_0$ that generates graphs around the observed sufficient statistic, the approximate MLE will not exist. However, this is difficult, since as the observed sufficient statistic approaches the boundary, the number of network configurations corresponding to it becomes smaller. This concept can be formalized by measuring the entropy, discussed next.

\paragraph{Entropy.} As explained in \cite{RinaldoFienbergZhu09} (see Section 3.4 therein for details),  the shape of the model polytope supports the argument that the full ERGM is ill-behaved. Specifically,  they use Shanon's entropy, which captures the degree to which the model concentrates its mass on network configurations associated with a relatively very small number of network statistics. The rationale is that degenerate models have large areas of low entropy.  The correspondence between the model polytope and model degeneracy derived by Rinaldo et al.\ shows that the extremal rays of the normal fan of the model polytope correspond to directions of the ridges of Shanon's entropy function where it converges to some fixed value.  These extremal rays are outer-normals of the facets (in our case, edges) of the polytope; we see that as $k$ grows, the polytope becomes `flatter' or, equivalently, the directions of the outer-normals of the edges on the lower hull get closer together, making the area of high entropy smaller. 
Although the exact plots are unavailable for the full ERGM on $n=18$, we know that for $n=9$ already the rays of normal fan concentrate in a small area of the plane implying that the model has low entropy and is degenerate for a vast majority of parameter values; cf.\  \cite[Figure 4A]{RinaldoFienbergZhu09}. 
As the authors in the said article justify, we use the mean value parameters to illustrate this behavior, where it can be clearly seen.

\begin{figure}[H]
	\centering 
		\begin{subfigure}[b]{.29\textwidth}
			\includegraphics[scale=0.25]{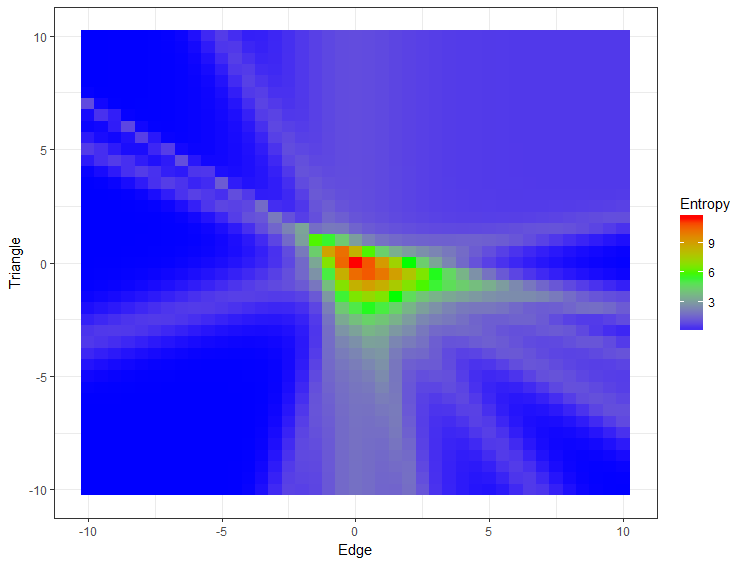}
			\subcaption{$n=18$, $k=3$}
		\end{subfigure}
\qquad
		\begin{subfigure}[b]{0.29\textwidth}
			\includegraphics[scale=0.25]{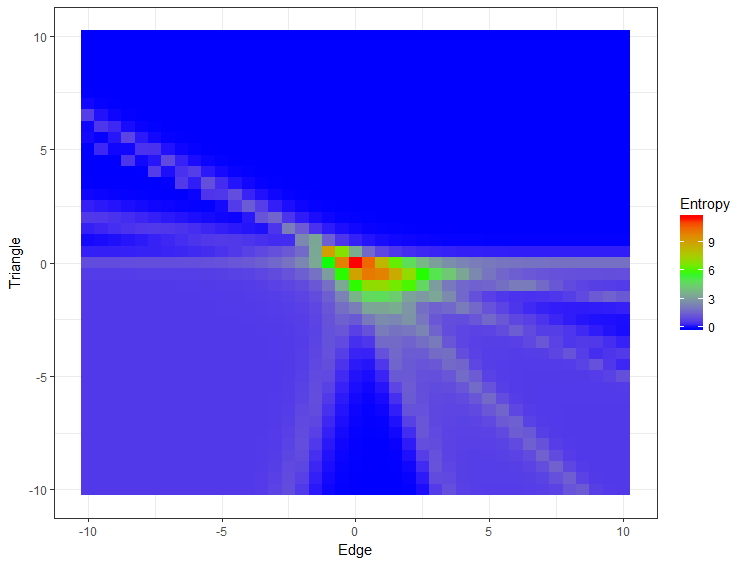}
			\subcaption{$n=18$, $k=4$}
		\end{subfigure}
\qquad
		\begin{subfigure}[b]{0.28\textwidth}
			\includegraphics[scale=0.25]{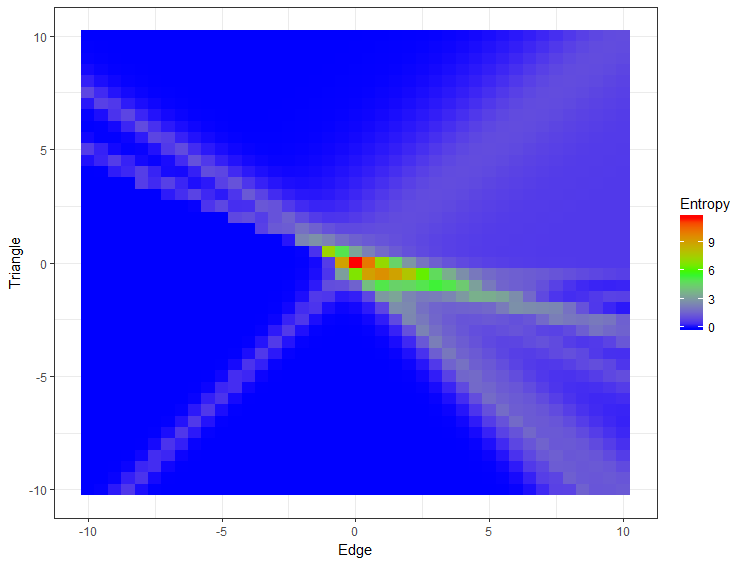}
			\subcaption{$n=18$, $k=5$}
		\end{subfigure}
	\caption{
		Comparison of degenerate (low-entropy) regions in the mean-value parameter space for the edge-triangle DERGMs on $n=18$ and increasing $k$. Sample sizes are $100,000$.  As $k$ increases, the high-entropy region becomes smaller. 
	}
	\label{fig:entropyFanRegions}
\end{figure}

In contrast, Figure~\ref{fig:entropyFanRegions} shows that the higher-entropy region is more `spread out' across the parameter region for the DERGMs with smaller values of $k$.  While one cannot, of course, conclude that the model is non-degenerate for all possible parameter values, it is clear that  the size of the parameter space that correspond to degenerate regions is  certainly less  than in the full ERGM. 
Regarding the caveat that the Figures  are also estimated and not exact, we are nevertheless confident in the results, because 1) the algorithm used is a uniform sampler of well-ordered graphs from the model support $\Gnk$; and 2) the estimated polytope is not far off from the true model polytope: it is missing some  extremal graphs that are probabilistically unlikely to be generated by the uniform sampler from the space of graphs $\G_9=\G_{9,8}$.

%%%%%%%%%%%%%%%%%%%%%%%%%%%%%%%%%%%%%%%

\subsection{The likelihood surface changes with $k$}
\label{sec:simulationsDegeneracy}

\begin{figure}[H]
\centering 
\begin{subfigure}[b]{.3\textwidth}
	\includegraphics[scale=0.2]{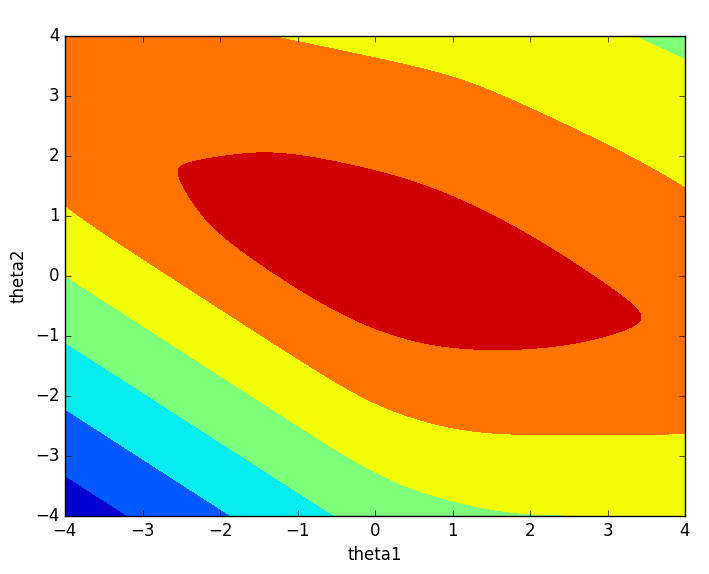}
	\subcaption{$n=18$, $k=3$}
\end{subfigure}
\quad
\begin{subfigure}[b]{.3\textwidth}
	\includegraphics[scale=0.2]{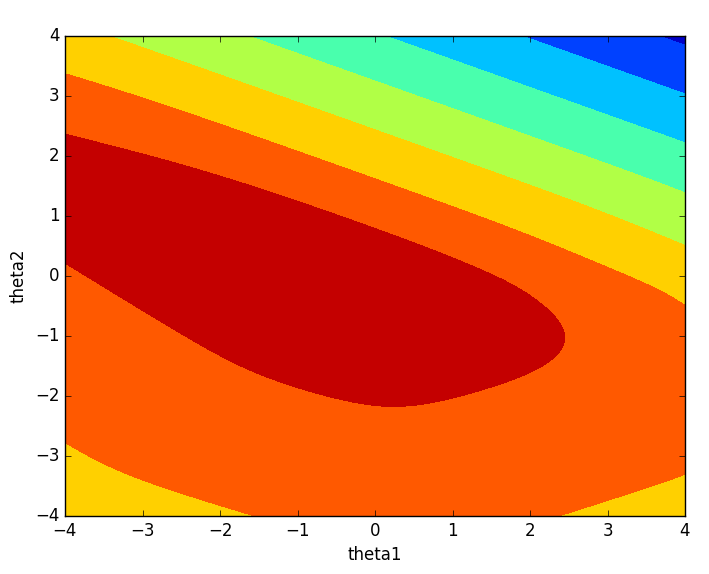}
	\subcaption{$n=18$, $k=4$}
\end{subfigure}
\quad
\begin{subfigure}[b]{.3\textwidth}
	\includegraphics[scale=0.2]{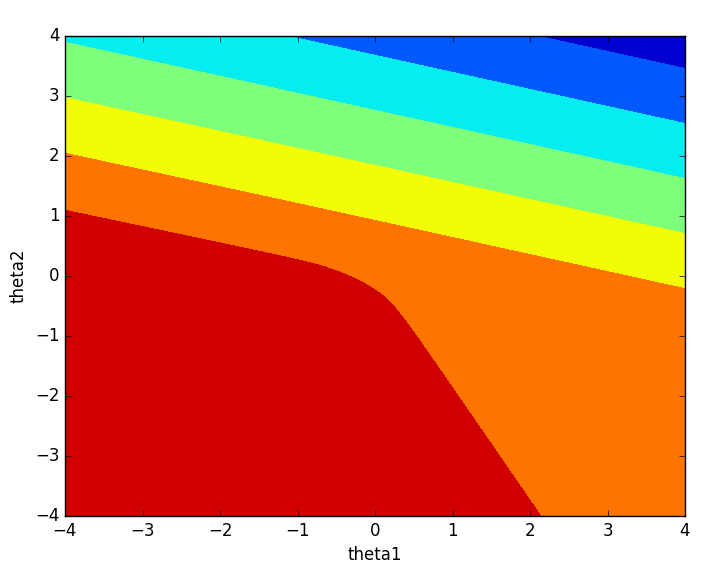}
	\subcaption{$n=18$, $k=17$}
\end{subfigure}
		\caption{
			Contour plots of the estimated edge-triangle DERGM likelihood functions for the Sampson network, for various values of $(\theta_1,\theta_2)$. Here, $n=18$ and $k=3,4,17$. Note that $k=17$ corresponds to the full ERGM. %\sonja{add (?): this was for a $200\times200$ grid.}  
			The estimated likelihood is based on an iid sample of $25,000$ graphs in $\G_{n,k}$. 
		}
		\label{fig:likelihoodCountours}
\end{figure}

The shape of the estimated likelihood function changes as we change $k$.  To illustrate this, we use the uniform sampler given in Algorithm~\ref{alg:uniformGnk} to sample graphs uniformly from the support of the full ERGM $\G_n=\G_{n,n-1}$ and $\Gnk$ with $k<n-1$ for various DERGMs, and estimate the likelihood function using the sampled graphs for the Sampson network. Figure~\ref{fig:likelihoodCountours} shows the contours of the (estimated) likelihood function for various values of $\theta = (\theta_{1},\theta_{2})$.
This figure uncovers an interesting trend:  the likelihood surface becomes `flatter' around the maximum value as $k$ grows, making it more difficult to find the maximum itself after a certain number of steps. 

\section{Estimation and fitting  DERGMs on real world data}\label{sec:simulationsFitting}

In this section, we present the results of fitting DERGMs to some real world networks. These results were obtained by fitting the DERGMs using the MCMC-MLE estimation algorithm using the tie-no-tie procedure, and \cite{hummel2012improving} step length algorithm to improve the estimation. The degeneracy parameter $k$ was set to its observed value. 

\subsection{Examples where DERGMs fit whereas ERGM fit fails to converge}
We first start by showing three examples where the MCMC-MLE procedure fails to converge when fitting an edge-triangle ERGM, whereas it converges when using the edge-triangle DERGM with the degeneracy parameter set to the observed degeneracy.  We consider three networks - an undirected version of the Sampson dataset, the \emph{Faux Mesa High} network and the undirected version of \emph{ecoli} network, from the {\tt ergm} package in {\tt R}. The summary statistics of these networks are given in Table \ref{table:datasets}. Note that we are not claiming that the edge-triangle DERGM is the best model for these data. Instead, the point is to illustrate that restricting the degeneracy has a direct impact on MCMC-MLE estimation.

\begin{table}
	\begin{center}
		\begin{tabular}{lc@{\hskip 0.1in}c@{\hskip 0.1in}c}
			\hline
			Network & Nodes  & Edges & Degeneracy \\
			\hline
			Sampson & 18 & 41 & 3 \\
			Faux Mesa High & 205 & 203 & 3 \\
			Ecoli & 418 & 519 & 3\\ 
			\hline
		\end{tabular}
		\caption{Summary of Datasets used to fit the edge-triangle DERGMs}
		\label{table:datasets}
	\end{center}
\end{table} 

The Sampson network has $n=18$ nodes and $m=41$ edges, with an observed degeneracy $k=3$. The Faux Mesa High network has $205$ nodes and $203$ edges, and an observed degeneracy of $3$. The ecoli network has $n=423$ nodes, $m=519$ edges with a degeneracy $k=3$. Note that all the networks have a low observed degeneracy. In particular, the ecoli and the faux mesa high networks are very sparse since the degeneracy is very small in comparison to the number of nodes. 

While fitting the edge-triangle ERGM to these networks, the MCMC-MLE combined with the step length procedure failed to converge due to model degeneracy; for a detailed study of this model's degeneracy, see \cite{RinaldoFienbergZhu09}. Specifically, the Markov chain started sampling networks whose number of edges and triangles are very far from the observed network, indicating model degeneracy. On the other hand,  there were no such issues when fitting the edge-triangle DERGM and the MCMC-MLE combined with the step length procedure converged. The estimated parameter for the edge-triangle DERGMs for these networks are given in Table \ref{table:dergmfits}.
There are two sources of standard error here, one from the MCMC estimation and another corresponding to the MCMCMLE. The MCMCMLE standard errors are calculated by using an MCMC estimate of the inverse of the estimated fisher information matrix, as described in \cite{HunterandHandcock}.

\begin{table}
	\begin{center}
		\begin{tabular}{l@{\hskip 0.1in} c@{\hskip 0.1in} c@{\hskip 0.1in} c}
			\hline
			Networks & Faux Mesa High & Sampson & Ecoli \\
			\hline
			edges          	& $-5.13^{***}$ & $-1.62^{***}$ & $-5.32^{***}$ \\
							& $(0.08)$      & $(0.34)$      & $(0.05)$\\
			triangle       	& $2.62^{***}$  & $0.36$        & $(2.65)^{***}$ \\
							& $(0.10)$      & $(0.34)$      & $(0.16) $\\
			\hline
			AIC            & 2029.17       & 157.41        & 6210\\
			BIC            & 2045.06       & 163.47        & 6229\\
			\hline
			\multicolumn{3}{l}{\scriptsize{$^{***}p<0.001$}}
		\end{tabular}
		\caption{Fitting the edge-triangle DERGM where the edge-triangle ERGM fit fails. The $^*$ denotes level of significance, based on the $p$-values. (The numbers in the parenthesis are the standard errors of the MCMCMLE.)
		}
		\label{table:dergmfits}
	\end{center}
\end{table}

\subsection{Examples when both ERGM and DERGM fit converges}
\label{sec:ergm&dergmfits}
We now consider cases where the MCMC-MLE procedure is able to fit both an ERGM and a DERGM to the same dataset. In these cases, we show that the parameter estimates obtained from both these models are very close to each other. We fit the edge-triangle DERGMs and ERGM to the \emph{florentine} dataset. This dataset has $n=16$ vertices and $m=20$ edges, with a degeneracy parameter $k=2$. We fit DERGMs with increasing values of $k=2,3,\ldots, 15$. Note that when $k=15$, the DERGM is equivalent to the edge-triangle ERGM. The parameter estimates are given in Table \ref{table:flo}. This table shows that the edge parameter is more or less the same for all the DERGMs and ERGM. The parameter corresponding to the triangles varies, but is within the margin of the standard error. 

\begin{table}
	\begin{center}
		\begin{tabular}{l c c c c c }
			\hline
			Degeneracy		 & 2 & 3 & 4 & 10 & 15 \\
			($k$)			 &     &    &   &    & (ERGM)\\
			\hline
			edges          	 & $-1.672^{***}$ & $-1.678^{***}$ & $-1.675^{***}$ & $-1.672^{***}$ & $-1.667^{***}$ \\
			& $(0.392)$      & $(0.362)$      & $(0.352)$      & $(0.346)$      & $(0.351)$          \\
			triangle       	 & $0.410$        & $0.172$        & $0.167$        & $0.152$        & $0.146$               \\
			& $(0.731)$      & $(0.595)$      & $(0.580)$      & $(0.572)$      & $(0.596)$      \\
			\hline
			AIC            & 111.786        & 112.058        & 112.073        & 112.090        & 112.071         \\
			BIC            & 117.361        & 117.633        & 117.648        & 117.665        & 117.646                \\
			Log Likelihood & -53.893        & -54.029        & -54.036        & -54.045        & -54.035                \\
			\hline
			\multicolumn{6}{l}{\scriptsize{$^{***}p<0.001$}}
		\end{tabular}
		\caption{Fitting DERGM and ERGM to the Florentine data. The $^*$ denotes level of significance, based on the $p$-values. (The numbers in the parenthesis are the standard errors of the MCMCMLE.)
		}
		\label{table:flo}
	\end{center}
\end{table}

%%%%%%=====================%%%%%%
\section{%The nuts and bolts of the underlying sampling algorithms or 
Uniform samplers for $\mathcal G_{n,k}$}
\label{sec:uniformSampler}
The main contribution of this section is the development of a fast uniform sampler of the space of well-ordered graphs in $\Gnk$, contained in Section~\ref{sec:uniformWO}, which has been used throughout  Section~\ref{sec:simulations} in simulations, most prominently for 
  estimated polytope plots. 
We discuss the basis of the algorithm and the updates we made to make it scalable.  
This algorithm can be used stand-alone for Monte Carlo sampling for DERGM estimation, specifically in the case when non-well-ordered graphs are not of interest. On the other hand, it can also be used in combination with a non-well-ordered sampler  to create a stratified sampler for all graphs of $\Gnk$ when needed; below, we discuss how in some cases the stratified sampler effectively reduces to the well-ordered one. Finally, if the observed graph is well outside the convex hull of sampled graphs, one may wish to use a fast importance MCMC sampler,  
 in conjunction with the uniform sampler from  Section~\ref{sec:uniformWO} to create an umbrella sampler on $\Gnk$.  The umbrella sampler converged quickly in simulations, but we omit those results here as they were not necessary for the data sets we analyze.

\subsection{A uniform sampler for well-ordered graphs from $\G_{n,k}$}\label{sec:uniformWO}
In \cite[Algorithm 1]{BauerEtAlSamplingDegenerate}, the authors derive a uniform sampler for the set of well-ordered graphs in $\G_{n,k}$. A \emph{well-ordered} graph is one in which the node labels are ordered so that no vertex has more than $k$ neighbors with a higher label. 

Using  this algorithm as a starting point, we make several key changes  to ensure that their algorithm is computationally efficient: we convert their  algorithm  from a recursive one to an iterative one. By doing this, we eliminate many complexity problems inherent in the original algorithm.  Specifically, the iterative version eliminates stack overflow issues for large graphs, as well as greatly reduces the execution time of generating a graph. 
 
Let us take a closer look at the following algorithm, based on  \cite[Algorithm 1]{BauerEtAlSamplingDegenerate}, which we   improved and updated to a scalable version. 

{\footnotesize  % OR: \scriptsize --these 2 are between \small and \tiny.
\begin{algorithm}[H]
\label{alg:uniformGnk}
\LinesNumbered
\DontPrintSemicolon
\SetAlgoLined
\SetKwInOut{Input}{input}
\SetKwInOut{Output}{output}
\Input{$n$, the number of nodes,\\
 $k$, maximum graph degeneracy.\\
}
\Output{$g$, a  graph in $\G_{n,k}$ in which every vertex $i$  has no more than $\geq k$ neighbors in the set $\{i+1,\dots,n\}$.}
\BlankLine
\For {$i=1$ to {$n$}}{$d_{i} \sim $ restrictedBinomial$(n-i,$ min$(n-i,k))$ \\
	\uIf {$i=n$}{$V = V \cup \{n\}$}}
\For {$i=n$ to 1}{$T = \{\}$ \\
	$P = V$ \\
	$a = |P|$ \\
	\For {$j=0$ to $d_{i}-1$}{ 
		$m \sim $ Uniform$(0,a-j)$ \\
		$T = T \cup \{(i, P_{m})\}$ \\
		$P_{m} = P_{a-j-1}$
		}
	$V = V \cup \{i\}$ \\
	$E = E \cup \{T\}$ }
$G = \{V, E\}$ \\
\Return G
\caption{Generate a well-ordered $g$ from $\G_{n,k}$ uniformly.}
\end{algorithm}
}

The algorithm was originally formulated using  recursion, which we emulate using two for-loops. The first for-loop populates a list of degrees where each index of the list corresponds to the respective vertex label. The degrees for each vertex are generated using a restricted binomial distribution. Instead of utilizing the cumulative distribution and using binary search to obtain values as suggested by the original paper, we opt to use the probability density function and store the values in a list data structure, reducing the complexity of obtaining the degree values. When the loop reaches the very last vertex, we add that vertex to the working vertex set. 
For each iteration in the second for-loop, a temporary copy of the current working vertex set is created. We then uniformly generate $d_{i}$ indices to sample without replacement from the vertex set copy, and use these samples for the edge set of the current vertex. It is obvious that this sample is uniformly generated, complying with the original algorithm.

For a benchmark, we tested the original recursive version (including generating all possible combinations) and the new iterative version on a machine with the following specifications: Intel Core i7-4790K CPU @ 4.00 GHz, 8 GB DDR3 RAM, Arch Linux x64, with the results shown in Table \ref{tab:runningtime}.  The results clearly indicate that the scalable version is superior in regards to time complexity.
%\medskip
\begin{table}
	\begin{center}
			\begin{tabular}{l@{\hskip 0.1in}l@{\hskip 0.1in}l}
			$(n,k)$	& Original Recursive Algorithm	&Our Iterative Version\\
					& \cite[Algorithm 1]{BauerEtAlSamplingDegenerate} & Algorithm~\ref{alg:uniformGnk}\\
			\hline
			$(50,8)$ 	&	$3.96$ seconds 		&	$0.03$ seconds\\
			$(800,2)$	& Stack Overflow		& 	$0.51$ seconds\\
			$(3000,2)$	& Stack Overflow		&	$1.90$ seconds\\
			\hline
		\end{tabular}
		\caption{Run times of the uniform samplers.}
	\label{tab:runningtime}
	\end{center}
\end{table}
In some applications, it may be desirable to further restrict the sample space of the model by restricting the total number of edges of the graph, or use such a restriction for stratified sampling of $\Gnk$. To that end, let $\G_{n,m,k}$ be the set of graphs on $n$ nodes and degeneracy $k$ with exactly $m$ edges. 
\cite[Algorithm 2]{BauerEtAlSamplingDegenerate} offer an algorithm for uniform sampling of $\G_{n,m,k}$, however, it was not implemented due to the complexity of step $3$ that the authors suggest be implemented using Equation (2.7) in \cite{BauerEtAlSamplingDegenerate}. Pre-computation of degrees proved nearly impossible in practice for several reasons. The recursive nature of calculating the cardinality for possible graphs of given vertices, edges, and degeneracy yielded very inefficient computations 
 in which the run time of each computation was longer than trying to generate whole graphs by other means. While we were able to alleviate this issue somewhat by utilizing a dynamic programming approach with memoization, even for semi-sparse, average size graphs, numerical overflow occurred, which rendered the speed increase fruitless. Instead, we opt to use \cite[Algorithm 3]{BauerEtAlSamplingDegenerate}, which is a non-uniform but fast sampler of $\G_{n,m,k}$.  Our implementation  of this algorithm, outlined in Algorithm~\ref{alg:nonuniformGnmk}, stays true to the pseudo-code given in the original paper, with the only alteration being utilizing the same approach to uniform selection as in our implementation of Algorithm~\ref{alg:uniformGnk}. 

{\footnotesize 
	\begin{algorithm}[H]
		\label{alg:nonuniformGnmk}
		\LinesNumbered
		\DontPrintSemicolon
		\SetAlgoLined
		\SetKwInOut{Input}{input}
		\SetKwInOut{Output}{output}
		\Input{$n$, the number of nodes,\\
			$m$, the number of edges,\\
			$k$, maximum graph degeneracy.\\
		}
		\Output{$g$, a  graph in $\in\G_{n,k}$  with $m$ edges in which every vertex $i$ has no more than $\geq k$ neighbors in the set $\{i+1,\dots,n\}$.}
		\BlankLine
		$C = {1,...,v_{n-1}}$\\
		\For {$i=1$ to {$m$}}{$j \sim $ Uniform$(0, |C|)$ \\
			$d_{j} = d_{j}+1$\\
			\uIf {$d_{j} = $ min$(n-v_{j},k)$}{$C$ \textbackslash $\{v_{j}\}$}}
		\For {$i=1$ to $n-1$}{$T = \{\}$ \\
			$P = V$ \\
			$a = |P|$ \\
			\For {$j=0$ to $d_{i}-1$}{ 
				$m \sim $ Uniform$(0,a-j)$ \\
				$T = T \cup \{(i, P_{m})\}$ \\
				$P_{m} = P_{a-j-1}$
			}
			$V = V \cup \{i\}$ \\
			$E = E \cup \{T\}$ }
		$G = \{V, E\}$ \\
		\Return G
		\caption{Generate a well-ordered $g$ from $\G_{n,m,k}$ non-uniformly.}
	\end{algorithm}
}

\subsection{Stratified sampling of $\Gnk$ to include non-well-ordered graphs if needed}\label{sec:stratified} 

Another issue with \cite[Algo.1]{BauerEtAlSamplingDegenerate} is that it generates only so-called `well-ordered' graphs in $\Gnk$.
 This misses a part of graphs in the support of our model. To remedy this issue, we classify all missing graphs and produce them via {stratified sampling} with two strata. Specifically, Algorithm~\ref{alg:uniformGnk} is used to sample from the set of well-ordered graphs in $\Gnk$, while Algorithm~\ref{alg:uniformGnkMissingGraphs}, described below, is used to generate non-well-ordered graphs in $\Gnk$. 
 Let $n_1$ and $n_2$ be the number of well-ordered and non-well-ordered graphs, respectively. The formula for $n_1$ is provided in \cite{BauerEtAlSamplingDegenerate} under the notation $D_n^{(k)}$, while $n_2$ is studied below. 
 To the best of our knowledge, the literature does not provide a good estimate of the number $n_1$ of well-ordered $k$-degenerate graphs compared to the total number of $k$-degenerate graphs. 
Although we derived a lower bound on the total number of $k$-degenerate graphs ($\Omega(n\log n)$) in Theorem~\ref{thm:lowerboundSupport},  in this section we study the ratio of $n_1$ and $n_2$ further, which is needed from an algorithmic point of view. 
It should be noted that, in practice, the uniform sampler from Section~\ref{sec:uniformWO} may only be omitting a tiny fraction of graphs in the support of the DERGM; this situation is described in detail at the end of this Section.  
 Therefore, the reader interested in applications more than in theory behind the algorithms that may not be necessary in practice  may skip the remainder of this technical section.  

\smallskip
A graph $g\in\Gnk$ is \emph{not well-ordered} if there exists at least one vertex $j$ with at least $k+1$ neighbors in the set $\{j+1,\dots,n\}$.  
Among all such vertices with too many big neighbors, let $k+c$ be the minimum such number of big neighbors, and let $i$ be the index of the smallest vertex that has $k+c$ big neighbors. 
We construct non-well-ordered graphs and use them to estimate $n_1$ by going through possible cases for the values of $c$ and $i$. 
For each case $c=1,\dots,n-k-1$, some vertex $i$  has $k+c$ neighbors in the set $\{i+1,\dots,n\}$. For each of the cases, the vertex $i$ can be chosen from the set $\{1,\dots,n-(k+c)\}$. Note that these $k+c$ neighbors of $i$ can be connected in any arbitrary way, as long as the entire graph is in $\Gnk$. 
Thus, we proceed as follows: construct a random graph $h$ on $k+c$ vertices whose labels are in the set $\{i+1,\dots,n\}$. Then, construct a suspension $g$ over $h$ using vertex $i$, that is, ensure that $i$ is connected to all $k+c$ vertices of $h$. 
Finally, the vertices $\{1,\dots,i\}$ can be connected in any way such that, by minimality of $i$, the resulting subgraph on $\{1,\dots,i\}$ is well-ordered and, additionally, each vertex in the set $\{1,\dots,i\}$ can have at most $k$ neighbors in the vertex set $\{i+1,\dots,n\}$. 
The construction is outlined in Algorithm~\ref{alg:uniformGnkMissingGraphs}. 

{\footnotesize  
\begin{algorithm}[H]
\label{alg:uniformGnkMissingGraphs}
\LinesNumbered
\DontPrintSemicolon
\SetAlgoLined
\SetKwInOut{Input}{input}
\SetKwInOut{Output}{output}
\Input{$n$, the number of nodes,\\
$k$, maximum graph degeneracy.\\
}
\Output{$g$, a  graph in $\in\Gnk$ (or $\G_{n,d}$ with $d>k$, unfortunately) in which there is a vertex $i$ that has $\geq k+1$ neighbors in the set $\{i+1,\dots,n\}$.}
\BlankLine
Pick $c\in\{1,\dots,n-k-1\}$. \label{line:start}\;% \sonja{$\leftarrow$ pick uniformly for now; discuss to get correct overall uniform later} \;
%\uIf {$c=1$}{Do one thing, consisting of many steps etc etc}
Pick $i\in\{1,\dots,n-(k+c)\}$.  \;% \sonja{$\leftarrow$ pick uniformly for now; discuss to get correct overall uniform later} \;
Use Algorithm~\ref{alg:uniformGnk} to sample $\tilde h\in\G_{k+c,k+c-1}$; repeat until $degen(\tilde h)\leq k$. \label{line:htilde}\; % \sonja{goes closer to eq(17).} 
Choose (uniformly) a subset of $k+c$ vertex labels from the set of legal vertex labels $\{i+1,\dots,n\}$. \label{line:vtx.shift}\;
Let $h$ be the graph obtained from $\tilde h$ by replacing the labels $1,\dots,k$ by those selected on Line~\ref{line:vtx.shift}. \label{line:h.by.shifting.htilde}\;
Create the suspension graph $g$ over $h$ by adding to $h$  edges $\{i,x\}$ for all $x\in V(h)$.  \label{line:g.suspension.from.h}\; 
Connect vertices $\{1,\dots,i\}$ by constructing any well-ordered graph from $\G_{i,k}$.\label{line:connect1thrui}\;
Connect any of the vertices $\{1,\dots,i\}$ to at most $k$ vertices in the set $\{i+1,\dots,n\}$.\label{line:connect1thruiToBig}\;
Output $g$ if $degen(g)\leq k$; otherwise return to Step~\ref{line:start}.   \;

\caption{Generate a non-well-ordered $g$ from $\Gnk$ 
}
\end{algorithm}
}
There are ${ n-i \choose k+c }$ ways to choose the neighbors of the vertex $i$ on Line~\ref{line:vtx.shift} and for each choice of neighbors there are $2^{k+c\choose 2}$ graphs $\tilde h$ generated on Line~\ref{line:htilde}. There are $D^{(k)}_i$ well-ordered graphs on Line~\ref{line:connect1thrui} and $i\sum_{p=1}^k{n-i \choose p}$ graphs on Line~\ref{line:connect1thruiToBig}. 
Thus, Algorithm~\ref{alg:uniformGnkMissingGraphs} constructs the following number of graphs $g$: 
\begin{align}  
	\underbrace{
	 %&
	  \sum_{i=1}^{n-(k+1)}\underbrace{{n-i\choose k+1}}_{\mbox{Line~\ref{line:vtx.shift}}}
		\cdot 
		\underbrace{2^{k+1\choose 2}}_{\mbox{Line~\ref{line:htilde}}}
		\cdot \underbrace{D_i^{(k)}}_{\mbox{Line~\ref{line:connect1thrui}}}
		\cdot \underbrace{i\sum_{p=1}^k{n-i \choose p}}_{\mbox{Line~\ref{line:connect1thruiToBig}}}
	}_{c=1} 
 \nonumber  \\ 
	 %&
	  		+ 
	\underbrace{
		\sum_{i=1}^{n-(k+2)}{n-i\choose k+2}\cdot 2^{k+2\choose 2} \cdot D_i^{(k)} \cdot i\sum_{p=1}^k{n-i \choose p} +\dots
	}_{c=2}
 \nonumber  \\ 
	 %&
	  	\dots +
	\underbrace{
	\sum_{i=1}^{n-(k+n-k-1)}{n-i\choose n-1}\cdot 2^{n-1\choose 2}\cdot D_i^{(k)} \cdot i\sum_{p=1}^k{n-i \choose p}}_{c=n-k-1}
	\\ 
	 %= &
	 =	2^{k+1\choose 2} \cdot\sum_{i=1}^{n-(k+1)}{n-i\choose k+1}\cdot  D_i^{(k)} \cdot i\sum_{p=1}^k{n-i \choose p}
 \nonumber  \\ 
	%& 
	        + 2^{k+2\choose 2} \cdot \sum_{i=1}^{n-(k+2)}{n-i\choose k+2}\cdot D_i^{(k)} \cdot i\sum_{p=1}^k{n-i \choose p}
	+\dots 
\nonumber \\
	%& 
	        \dots + 	2^{n-1\choose 2}\cdot  {n-1\choose n-1}\cdot D_i^{(k)} \cdot i\sum_{p=1}^k{n-i \choose p},
\label{eq:NumGraphsFromNWOalgorithm}  
\end{align}
where each of the $n-k-1$ summands corresponds to one of the cases $c$.  

Note that Equation~\eqref{eq:NumGraphsFromNWOalgorithm} is  an upper bound on $n_2$, since it counts all graphs $g$ constructed by Algorithm~\ref{alg:uniformGnkMissingGraphs}. 
It is also a strict upper bound on the number of graphs $g$ actually returned by the algorithm, since it 
counts those graphs whose degeneracy happens to be strictly larger than $k$.

 Equation~\eqref{eq:NumGraphsFromNWOalgorithm} counts all graphs on $k+c$ nodes, $2^{k+c\choose 2}$, constructed in Step~\ref{line:htilde}. Surely, a better count can be obtained by replacing $2^{k+c\choose 2}$ by $$2^{k+c\choose 2}-\#\{\mbox{well-ordered graphs on $k+c$ vertices of degeneracy}>k\}.$$ 
Doing this replacement in the equation is, crucially, still an upper bound on $n_2$ (since the well-ordered graphs of degeneracy larger than $k$ certainly do not contribute to any non-well-ordered graphs of degeneracy at most $k$). 
Since
\begin{align*}
	   &\#\{\mbox{well-ordered graphs on $k+c$ nodes of degeneracy}>k\}  \\
	=&\#\{\mbox{all well-ordered graphs on $k+c$ nodes except those of degereacy} \leq k\}  \\ 
	=& D^{(k+c-1)}_{k+c}-D^{(k)}_{k+c},  %this is obvious: all graphs of degen at most $k+c-1$, which is all graphs on $k+c$ nodes, 
\end{align*} 
the following is a better upper bound on the number of graphs we wish to keep from Algorithm~\ref{alg:uniformGnkMissingGraphs} and thus also an upper bound on  $n_2$: 
\begin{align}\label{eq:BETTERNumGraphsFromNWOalgorithmWeAreKeeping}
	\sum_{i=1}^{n-(k+1)}{n-i\choose k+1} &\cdot  \left( 2^{k+1\choose 2}-\left( D^{(k+1-1)}_{k+1}-D^{(k)}_{k+1}\right)\right) \cdot  D_i^{(k)} \cdot i\sum_{p=1}^k{n-i \choose p} +
	 	\nonumber \\
	 \sum_{i=1}^{n-(k+2)}{n-i\choose k+2} &\cdot   \left( 2^{k+2\choose 2} -\left( D^{(k+2-1)}_{k+2}-D^{(k)}_{k+2}\right)\right) \cdot   D_i^{(k)} \cdot i\sum_{p=1}^k{n-i \choose p}
	+\dots \nonumber \\
	\dots + 	 {n-1\choose n-1} &\cdot   \underbrace{\left( 2^{n-1\choose 2} -\left(D^{(n-1-1)}_{n-1}-D^{(k)}_{n-1}\right)\right) }_{\mbox{Line~\ref{line:htilde} minus well-ordered  of degen}>k}\cdot   D_i^{(k)} \cdot i\sum_{p=1}^k{n-i \choose p}. 
\end{align} 

Let
$$t_{true}=\log n_1/(n_1+n_2)$$ 
be the true threshold used to divide the sample in two strata and define 
$$t_{estimated}=\log n_1/(n_1+\eqref{eq:BETTERNumGraphsFromNWOalgorithmWeAreKeeping}).$$ 
Given that $\eqref{eq:BETTERNumGraphsFromNWOalgorithmWeAreKeeping}>n_2$,  % $c>0$, 
 $ t_{estimated} < t_{true}\leq 0$.  
Therefore we take the following approach: 
1) compute the threshold $t_{estimated}$ for the fixed n and k for which we wish to run the current simulation.
2) If $t_{estimated}$ is close to $0$, then that forces $t_{true}$ to be close to $0$, which in turn means that there is a very, very small number of non-well-ordered graphs for that choice of $n$ and $k$ and therefore the stratified sampler essentially reduces to sampling well-ordered graphs only. 

Of course,  if $t_{estimated}$ is not  relatively close to $0$, then  for those values of $n$ and $k$, while it is possible that $t_{true}$ is close to $0$, one should implement both the well-ordered and non-well-ordered algorithm. 
Falling back on the well-ordered algorithm is equivalent to using an approximate sampler in practice. The users may additionally prefer to replace Algorithm~\ref{alg:uniformGnkMissingGraphs} by instead permuting the vertices of the output of Algorithm~\ref{alg:uniformGnk}, allowing it to reach the entire sample space $\Gnk$ in another way. 

\begin{remark} \rm
In practice, if the model's sufficient statistics are subgraph counts (or if the distribution is exchangeable), well-ordering does not pose a restriction, because in the uniform sampling using MC in estimating the MLE,  only the values of the sufficient statistics of the sampled graphs are used. These are oblivious to vertex labels, so ordering is irrelevant. 
\end{remark}

%%%%%%=====================%%%%%%

\section{Discussion}
\label{sec:discussion}

In this paper, we introduced a general modification of exponential family random graph models that solves some of the model degeneracy issues. This modification amounts to a support restriction, by conditioning on the observed network's graph-degeneracy, which is a measure of sparsity that is weaker than imposing an upper bound on node degrees. The resulting model class, which we name \emph{degeneracy-restricted} or \emph{DERGMs}, does not suffer from the same estimation issues as the usual ERGMs. 
The proposed support restriction is interpretable as a weak sparsity constraint, it respects most real-world network data, and it provably does not eliminate a large part of the support of the full ERGM, while improving model behavior.  Specifically, we show that DERGMs with smaller graph degeneracy parameter $k$ induce stable sufficient statistics, and we also show that such a stable behavior implies non-degeneracy of the model. Using simulations, we also show that DERGMs  with small values of $k$ have a better-behaved simulated likelihood (i.e., more steep around the maximum) and the simulated model polytope spreads more mass around realistic graphs by eliminating very low-probability extreme graphs. This also makes MCMC algorithms to approximate the likelihood more stable, thus improving the MCMC-MLE estimation.

The particular example of the edge-triangle DERGM presented here is a good illustration of the general DERGM behavior. It is a natural choice of the running example, given the recent work by \cite{RinaldoFienbergZhu09} that studies its degenerate behavior in detail. The general framework presented, however, applies to any ERGM; a good overview of many of the popular classes being offered in  \cite{F-review}. 
Recent work on the shell-distribution ERGM  \cite{KarPelPetStaWilb:shellErgm} introduces a limited version of the current contribution: it is an example of an ERGM with similarly restricted support and gives direct motivation for the study of DERGMs in general. However, there, the model support was not $\Gnk$ for fixed $n$ and $k$, but rather $\Gnk\setminus \G_{n,k-1}$ - networks with degeneracy exactly $k$. Here were propose to use networks of degeneracy at most $k$, to enlarge the model support, and offer greater flexibility in modeling.  
Our contributions indicate that  DERGMs may offer a feasible and interpretable modification of ERGMs, a powerful and flexible model class.

Extending the approach presented herein to directed graphs is one of the directions of future work. The notion of $k$-degeneracy as defined here applies only to undirected graphs, however it has been extended to directed graphs recently in \cite{D-core}. 
Another direction of future is to develop  a distributed version of Algorithm~\ref{alg:uniformGnk}. While we did run the current implementation in parallel, it can further be improved to run on a cluster. The current implementation scales very well to hundreds of nodes and with the additional step it should perform just as well on thousands.

\bibliography{degERGMs}
%\bibliography{ref,AlgStatNtwks,kCoresERGM}

\end{document}